\documentclass[11pt,a4paper]{article}
\usepackage{authblk}
\usepackage[english]{babel}
\usepackage[T1]{fontenc}
\usepackage[utf8]{inputenc}
\usepackage{amsmath,amsfonts,amsthm,amssymb}
\usepackage{comment}
\usepackage{mathtools}
\usepackage{bbold}
\usepackage{palatino}
\usepackage{eurosym}
\usepackage{enumerate}
\usepackage[symbol]{footmisc}

\usepackage[toc,page]{appendix}
\usepackage[inline]{enumitem}
\usepackage[%
    colorlinks=true,
    pdfborder={0 0 0},
    linkcolor=red
]{hyperref}
\usepackage{hyperref}
\hypersetup{colorlinks,linkcolor={blue},citecolor={blue},urlcolor={red}}  
\usepackage[top=2.5cm, left=2.5cm, right=2.5cm, bottom=2.5cm]{geometry}
\usepackage[square,numbers]{natbib}
\numberwithin{equation}{section}
\title{A mathematical model to study the impact of intra-tumour heterogeneity on anti-tumour CD8$^+$ T cell immune response\\
}
\date{\vspace{-6ex}}
\vspace{3ex}
\author[1]{Emma Leschiera\thanks{Corresponding author \\
\textit{Email addresses}: \texttt{leschiera@sorbonne-universite.fr} (Emma Leschiera), \texttt{tommaso.lorenzi@polito.it} (Tommaso Lorenzi),
\texttt{shenshensi@wchscu.cn} (Shensi Shen), \texttt{almeida@sorbonne-universite.fr} (Luis Almeida), \texttt{chloe.audebert@sorbonne-universite.fr} (Chloe Audebert).\\
E.L. has received funding from the European Research Council (ERC) under the European Union’s Horizon2020 research and innovation programme (grant agreement No 740623). \\
T.L. gratefully acknowledges support of the MIUR grant ``Dipartimenti di Eccellenza 2018-2022''.}}
\author[2]{Tommaso Lorenzi}                                     \author[3]{Shensi Shen}
\author[1]{Luis Almeida}
\author[5,6]{Chloe Audebert}

\affil[1]{\small\textit{Sorbonne Universit\'e, CNRS, Universit\'e de Paris, Inria, Laboratoire Jacques-Louis Lions UMR 7598, 75005 Paris, France.}}
\affil[2]{\textit{Department of Mathematical Sciences ``G. L. Lagrange'', Dipartimento di Eccellenza 2018-2022, Politecnico di Torino, 10129 Torino, Italy.}}
\affil[3]{\textit{Institute of Thoracic Oncology, West China Hospital, Sichuan University, Chengdu, China.}}
\affil[4]{\textit{Sorbonne Universit\'e, CNRS, Universit\'e de Paris, Laboratoire Jacques-Louis Lions UMR 7598, 75005 Paris, France.}}

\affil[5]{\textit{Sorbonne Universit\'e, CNRS, Institut de biologie Paris-Seine (IBPS), Laboratoire de Biologie Computationnelle et Quantitative UMR 7238, 75005 Paris, France.}}

\begin{document}
\maketitle

\begin{abstract}
Intra-tumour heterogeneity (ITH) has a strong impact on the efficacy of the immune response against solid tumours. The number of sub-populations of cancer cells expressing different antigens and the percentage of immunogenic cells (\textit{i.e.} tumour cells that are effectively targeted by immune cells) in a tumour are both expressions of ITH. Here, we present a spatially explicit stochastic individual-based model of the interaction dynamics between tumour cells and CD8$^+$ T cells, which makes it possible to dissect out the specific impact of these two expressions of ITH on anti-tumour immune response. The set-up of numerical simulations of the model is defined so as to mimic scenarios considered in previous experimental studies. Moreover, the ability of the model to qualitatively reproduce experimental observations of successful and unsuccessful immune surveillance is demonstrated. First, the results of numerical simulations of this model indicate that the presence of a larger number of sub-populations of tumour cells that express different antigens is associated with a reduced ability of CD8$^+$ T cells to mount an effective anti-tumour immune response. Secondly, the presence of a larger percentage of tumour cells that are not effectively targeted by CD8$^+$ T cells may reduce the effectiveness of anti-tumour immunity. Ultimately, the mathematical model presented in this paper may provide a framework to help biologists and clinicians to better understand the mechanisms that are responsible for the emergence of different outcomes of immunotherapy.

\end{abstract}


\vspace{5ex}


\section{Introduction} \label{Introduction}

Recent  technological advances have allowed for the design of immunotherapy which, in contrast to conventional anti-cancer therapies, targets tumour-immune cell interactions with the aim of re-boosting the effectiveness of the anti-tumoural immune responses. Although immunotherapy has revolutionized anti-tumour treatment, its efficacy remains limited in most clinical settings  \citep{anagnostou2017evolution,basu2016cytotoxic,boissonnas2007vivo,galon2019approaches,ribas2018cancer,topalian2015immune}. \\
Immune cells, specifically CD8$^+$ cytotoxic T cells, are capable of detecting and eliminating tumour cells by recognising cancer-associated antigens expressed by tumour cells. The effectiveness of the immune response depends on the level of presentation of such antigens by the major histocompatibility complex 1 (MHC-I) \citep{coulie2014tumour,messerschmidt2016cancers}. In particular, CD8$^+$ T cells express a unique repertoire of T cell receptors (TCRs) \citep{parkin2001overview} and, once activated, they migrate via chemotaxis in response to concentration gradients of chemical signals toward the tumour cells expressing the matching antigens \citep{miller2003autonomous}. The influx and movement of CD8$^+$ T cells are dictated by the spatial distribution of tumour antigens and by the level of chemokines in the tumour micro-environment \citep{boissonnas2007vivo}. Upon intratumoural infiltration,  CD8$^+$ T cells can trigger tumour cell death by direct interaction with tumour cells, releasing cytotoxic factors (\textit{i.e.} granzime B, interferon gamma) \citep{kim2007cancer}.\\

Oncogenic mutation-driven cancers harbor neoantigens that can be recognized by CD8$^+$ T cell receptors \citep{gubin2014checkpoint}. A high mutational burden and neoantigen load in tumours have been associated with an enhanced response to immunotherapy \citep{chan2019development,germano2017inactivation,hellmann2018genomic,samstein2019tumor,snyder2014genetic,van2015genomic}. However, it has recently been reported that many of these neoantigens arise from sub-clonal branching mutations and could potentially increase intratumour heterogeneity (ITH) \citep{mcdonald2019tumor,mcgranahan2016clonal,reuben2017tcr}. 
These tumours are characterised by clonal antigens (presented by all tumour cells), and sub-clonal antigens (presented only by sub-populations of tumour cells).
Moreover, such sub-clonal antigens may be associated with decreased level of antigen presentation by the MHC-I, leading to a weaker antigen-specific CD8$^+$ T cell response \citep{gejman2018rejection}. In contrast, more homogeneous tumours express few clonal antigens in all tumour cells and appear to have a better response to immunotherapy across a wide range of tumour types \citep{fennemann2019attacking,mcgranahan2016clonal}. Furthermore, CD8$^+$ T cells activated against clonal antigens are more commonly found at the tumour site than CD8$^+$ T cells reactive to sub-clonal antigens \citep{mcgranahan2016clonal}. These findings suggest that ITH may strongly affect the effectiveness of the anti-tumour immune response. \\ 

Mathematical models are useful tools for simulating and investigating biological systems, and have been increasingly used to investigate the role of tumour antigens and the effect of ITH on the anti-tumour immune response. Tumour heterogeneity and the role of tumour antigens have been studied using differential equation models \citep{aguade2020tumour,asatryan2016evolution,balachandran2017identification,lorenzi2016tracking} and cellular-automaton (CA) models \citep{bouchnita2017hybrid,ghaffarizadeh2018physicell}. A number of mathematical models have also been developed to investigate the dynamics of tumour development in the presence of adaptive immune response. Usually, these models are formulated as either ordinary differential equations \citep{de2003mathematical,kirschner1998modeling,kuznetsov1994nonlinear,luksza2017neoantigen} or integro-differential equations \citep{delitala2013recognition,kolev2013numerical,lorenzi2015mathematical}. Most of these models rely on the assumption that cells are well-mixed and, as such, do not take into account spatial dynamics of immune cells and tumour cells. Spatial and temporal dynamics of tumour-immune competition have been studied through partial differential equation (PDE) models \citep{atsou2020size,matzavinos2004travelling,matzavinos2004mathematical}. However, differential equation models are defined on the basis of population-level phenomenological assumptions, which may limit the level of biological detail that can be included in the model. By using computational models, such as CA and individual-based models (IBMs), a more direct and precise mathematical representation of biological phenomena can be achieved. These models can be posed on a spatial domain (\textit{e.g.} a grid), where cells interact locally with each other according to a defined set of probabilistic rules, and can collectively generate global emergent behaviours of tumour-immune cell interactions. A number of IBMs \citep{christophe2015biased,kather2017silico,macfarlane2018modelling} and hybrid PDE-CA models \citep{de2006spatial,kim2012modeling,mallet2006cellular} have also been used to study the interaction dynamics between tumour and immune cells. These models take into account different aspects of the anti-tumoural immune response (\textit{e.g.} expression of immunosuppressive factors, movement of immune cells) and clarify the conditions for the emergence of a range of situations of successful and failed immune response. However, they do not take into account the effects of antigen presentation and ITH on immune surveillance.
\\

In light of these considerations, we present a spatial stochastic individual-based model of tumour-immune interaction dynamics that can be used to explore the
effect of ITH on immune surveillance. There is a variety of individual-based model approaches (\emph{e.g.}, cellular automata, Cellular Potts models, hybrid discrete/continuous models). In our study, we used a Cellular Potts model and the CompuCell3D open-source simulation environment \citep{izaguirre2004compucell}. The originality of this model lies in the characterisation of antigen presentation levels by tumour cells, which drive the influx of CD8$^+$ T cells in the tumour micro-environment and their movement toward tumour cells. In our model, the effectiveness of the anti-tumour immune response is directly linked to the level of presentation of tumour antigens. In addition, the model takes into account biological phenomena that are driven by stochastic aspects of the interaction dynamics between tumour cells and CD8$^+$ T cells. 
The effect of ITH on immune surveillance is investigated at two different levels through computational simulations of this model. 
First, we explore the outcomes of the immune response considering different number of sub-populations of cancer cells constituting the tumour. Then, we asses the efficiency of the immune response by varying the immunogenicity of tumour cells. We study the impact of these two characteristics on tumour progression independently and together, assessing their influence on anti-tumour immunity in a controlled manner.\\


The paper is organised as follows. In Section \ref{Models and methods}, we present the individual-based model and the mathematical description of each biological process included in the model. Section \ref{Computational simulations} summarises the set-up of computational simulations and presents some preliminary results of computational simulations. Full details of model implementation and model parametrisation are provided in \ref{Details of computational model} and \ref{appendix:graph}, respectively. In Section \ref{Main results} we present the main computational results and we discuss them in view of previous biological works. Finally, Section \ref{Discussion and conclusion} concludes the paper and provides a brief overview of possible research perspectives.

\section{Model and methods} \label{Models and methods}

We consider two cell types in our model: tumour cells, characterised by an antigen profile and a level of antigen presentation, and antigen-specific CD8$^+$ T cells. To describe the interactions occurring between the two cell types we use an on-lattice individual-based model posed on a 2D spatial domain partitioned into square elements of side $\Delta x$. In our model, this domain biologically represents the tumour micro-environment. At each time step of length $\Delta t$, the states of the cells are updated according to the probabilistic and deterministic rules described
below. \\
In the remainder of this section, we first present the modelling framework in a general setting, along with the underlying biological hypotheses and assumptions. Then, we detail how each biological mechanism is mathematically described. A detailed description of the computational implementation of the model, which relies on a Cellular Potts approach, can be found in \ref{Details of computational model}.

\subsection{Modelling framework}
To include different level of immunogenicity in the tumour, two different subtypes of tumour cells are considered: immunogenic cells (\textit{i.e.} tumour cells that are effectively targeted by CD8$^+$ T cells) and non-immunogenic cells (\textit{i.e.} tumour cells that are poorly targeted by CD8$^+$ T cells). On the one hand, we define immunogenic cells as cells expressing one or more clonal antigens, considered as immunodominant, and presented at a normal level by the MHC-I. On the other hand, we assume that non-immunogenic cells have experienced, through mutations, a deterioration of their level of antigen presentation, and have acquired new antigens. These new antigens are presented only by a subset of tumour cells, and will be denoted as sub-clonal antigens \citep{mcgranahan2016clonal}. Therefore, we define non-immunogenic cells as cells expressing clonal and sub-clonal antigens, both presented at a low level by the MHC-I. \\
The system is initially composed of tumour cells only, which grow and proliferate through mitosis. Tumour cells secrete different chemoattractants that trigger the influx of specific CD8$^+$ T cells into the domain. When they arrive in the domain, CD8$^+$ T cells move via chemotaxis toward tumour cells expressing the matching antigens and, upon contact, try to eliminate them. \\
The modelling strategies used to reproduce these dynamics are described in detail in the following subsections, and are also schematically illustrated with an example in Figure \ref{fig:resumeHp} and Figure \ref{fig:resumeHp_2}.
\subsubsection{Dynamics of tumour cells} \label{Tumour cells}
\begin{figure}
    \centering
    \includegraphics[scale=0.42]{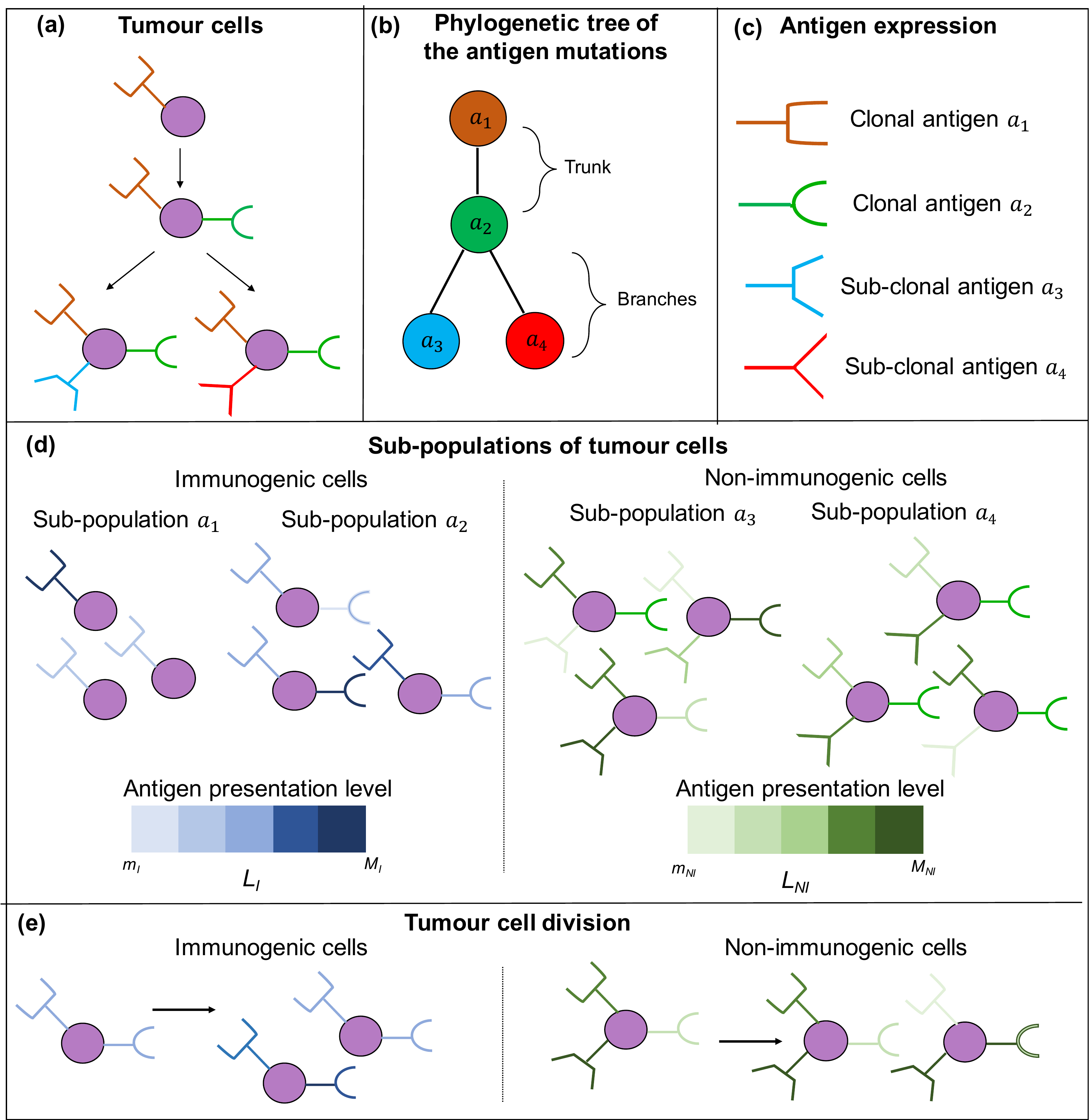}
    \caption{\textbf{Schematic representation of the modelling assumptions for tumour cells.} \textbf{(a)} Purple circles represent tumour cells. In this example, the antigen profile of the tumour is characterised by 4 different antigens, each one represented by a specific color and shape. \textbf{(b)} Phylogenetic tree illustrating the mutations leading to the different antigens expressed by tumour cells. The clonal and sub-clonal antigens are represented as the phylogenetic tree trunk and branches, respectively. \textbf{(c)} In this example, 4 antigens are expressed by the tumour, each one characterised by a different color and shape. Based on the phylogenetic tree \textbf{(b)}, we denote $a_1$ and $a_2$ as clonal antigens, whereas $a_3$ and $a_4$ are denoted as sub-clonal antigens. \textbf{(d)} The tumour is divided in 4 sub-populations of tumour cells, labelled by the last antigen acquired by each cell. Here, the color of each antigen represents its level of antigen presentation. Cells in the sub-populations labelled by the antigens $a_1$ and $a_2$ express only clonal antigens and are defined as immunogenic cells. They present their antigens at a normal level, with values chosen from the discrete set $L_{I}=\{\text{m}_I,\dots,\text{M}_I\}$. Cells in the sub-populations labelled by the antigens $a_3$ and $a_4$ express clonal and sub-clonal antigens and are defined as non-immunogenic cells. They present all their antigens at a low level, with values chosen from the discrete set $L_{NI}=\{\text{m}_{NI},\dots,\text{M}_{NI}\}$. \textbf{(e)} A tumour cell divides when it reaches a certain target volume. An immunogenic (respectively non-immunogenic) cell divide in two immunogenic (respectively non-immunogenic) cells. The daughter cell has the same antigens of the parent cell, but with a new random level of antigen presentation.}
    \label{fig:resumeHp}
\end{figure}

\paragraph{Antigen expression}
We let $N_T(t)$ denote the number of tumour cells in the system at time $t=h\Delta t$, with $h\in \mathbb{N_0}$, and we label each cell by an index $n=1,\dots, N_T(t)$. We let each tumour cell express one or more antigens, and we characterise the antigen profile of the tumour by means of a vector 
\begin{equation}
    A=(a_1,\dots,a_f), \quad a_1,\ldots, a_f \in \mathbb{N},
\end{equation}
where $a_i$ denotes an antigen and $f$ is the total number of antigens expressed by the tumour [see Figure \ref{fig:resumeHp}(a)]. Using phylogenetic tree representations [see Figure \ref{fig:resumeHp}(b)-(c)], we define each antigen $a_{i}\in A$, $i=1,\dots, f$, of the tumour as clonal if it belongs to the trunk of the phylogenetic tree, or sub-clonal if it belongs to one of the branches of the phylogenetic tree. We let $A_{C}$ and $A_{SC}$ denote the sets of clonal and the sub-clonal antigens, whereby:
\begin{equation}
  A_{C},A_{SC} \subset A, \quad  A_{C}\cup A_{SC}=A \quad \text{and} \quad A_{C}\cap A_{SC}= \emptyset.
\end{equation}
Then, based on the phylogenetic tree representation, we divide the tumour in $f$ different sub-populations of tumour cells labelled by the last antigen $a_{i}\in A$ acquired [see Figure \ref{fig:resumeHp}(d)]. Following this notation, in this model, cells in the same sub-population express the same set of antigens [see Figure \ref{fig:resumeHp}(a, b, d)]. Moreover, if $a_{i}\in A_{C}$, cells in the sub-population labelled by the antigen $a_{i}$ express only clonal antigens, whereas if $a_{i}\in A_{SC}$, cells in the sub-population labelled by the antigen $a_{i}$ express both clonal and sub-clonal antigens. Therefore, we define cells in sub-populations labelled by a clonal antigen $a_{i}\in A_{C}$ as immunogenic cells, whereas cells in sub-populations labelled by a sub-clonal antigen $a_{i}\in A_{SC}$ are defined as non-immunogenic cells. 

\paragraph{Antigen presentation by MHC-I}
We incorporate antigen presentation into our model by letting each tumour cell present its antigens at a certain level. There can be high variability in each antigen’s presentation between patients with the same type of tumour and even within tumour cell samples from the same patient \citep{anagnostou2017evolution,mcgranahan2016clonal}. Therefore, for the $n^{th}$ tumour cell, we characterise the level of presentation of each one of its antigens $a_i\in A$ by the normalized variable
\begin{equation}
    l_{a_i}^n \in \left[0,1\right]
\end{equation}
whereby the value $l_{a_i}^n=0$ corresponds to a tumour cell that lost the expression of the antigen $a_i$, while $l_{a_i}^n=1$ corresponds to a tumour cell presenting the antigen $a_i$ at the highest level.\\
To capture the idea that immunogenic cells present their antigens at a higher level than non-immunogenic cells, we introduce the discrete sets 
\begin{equation}
     L_{I}=\{\text{m}_I,\dots,\text{M}_I\}\subset [0,1] \; \text{and} \; L_{NI}=\{\text{m}_{NI},\dots,\text{M}_{NI}\}\subset [0,1], \; \text{with}  \; \text{M}_{NI}<\text{M}_{I}.
\end{equation}
They characterise the range of different values that can be taken by the variable $l_{a_i}^n$ [see Figure \ref{fig:resumeHp}(d)]. In particular, if the $n^{th}$ tumour cell is an immunogenic cell, it presents each antigen $a_i$ at a normal level $l_{a_i}^n \in L_{I}$, whereas if the $n^{th}$ tumour cell is a non-immunogenic cell, all of its antigens $a_i$ are presented at a low level $l_{a_i}^n \in L_{NI}$.
\paragraph{Tumour cell growth and division}
At each time-step, we let tumour cells grow at a random rate drawn from a uniform distribution; the parameters of the bounds of the uniform distribution are chosen to match the mean duration of a tumour cell cycle length. 
Mitosis occurs when a tumour cell grows to a critical size and then divides along a randomly orientated axis. Upon division at the time $t$, the $n^{th}$ tumour cell is replaced by two cells [see Figure \ref{fig:resumeHp}(e)], one labelled by the parent index $n$ and the other one labelled
by the index $N_T(t)+1$. 
The daughter cell will inherit most of the properties of the parent cell, including the antigens expressed by the parent cell, so the fact that the cell is immunogenic or not [see Figure \ref{fig:resumeHp}(e)]. For each antigen $a_i$ expressed by the daughter cell, a random level of antigen presentation $l_{a_i}
^{N_T(t)+1}$ will be chosen.This level of antigen presentation can then be different from the one of the parent cell. Another property not inherited by the daughter cell is the intrinsic lifespan of the cell, which is randomly drawn from a uniform distribution. In this model, we do not take into account the appearance of new antigens due to the occurrence of mutations.

\paragraph{Tumour cell death}
If a tumour cell exhausts its lifespan (which is drawn when the cell is created), it dies (\textit{i.e.} it undergoes apoptosis) at the end of the time-step and it is removed from the domain. 
A tumour cell can also die due to intra-tumour competition, with a rate proportional to the total number of tumour cells, or because of the cytotoxic action of CD8$^+$ T cells. More details about tumour cell death due to the cytotoxic action of CD8$^+$ T cells will be given in Section \ref{Dynamics of CD8$^+$ T cells}.
\paragraph{Secretion of chemoattractants}
We let tumour cells at the border of the tumour (the region where cytokines and immune cells are more abundant \citep{boissonnas2007vivo}) secrete different chemoattractants for each expressed antigen $a_i\in A$. The secretion of a chemoattractant by a tumour cell expressing antigen $a_i$ is proportional to the level of presentation of such antigen $a_i$. Therefore, we model the chemoattractant secretion rate $s_{a_i}^n$ by the $n^{th}$ tumour cell expressing antigen $a_i$ using the following definition:
\begin{equation}
s_{a_i}^n:=C_1\, l_{a_i}^n,
\label{antigen_pres}
\end{equation}
where $C_1 \in \mathbb{R^+}$ is a scaling factor of units $\frac{[\textit{mol}]}{[\textit{time}]\, [\textit{space}]}$, where $[\textit{mol}]$, $[\textit{time}]$ and $[\textit{space}]$ denote respectively the number of chemoattractant molecules and the units of time and of the size of a grid site, and $l_{a_i}^n$ is the level of presentation of antigen $a_i$ by the $n^{th}$ tumour cell.\\
The total amount of chemoattractant secreted by tumour cells expressing antigen $a_i$ induces the arrival of CD8$^+$ T cells specific to antigen $a_i$ into the domain. More details about the mathematical modelling of the different chemoattractant dynamics will be discussed in Section \ref{chemoattractant field}. 
\subsubsection{Dynamics of CD8$^+$ T cells}
\label{Dynamics of CD8$^+$ T cells}
\begin{figure}
    \centering
    \includegraphics[scale=0.42]{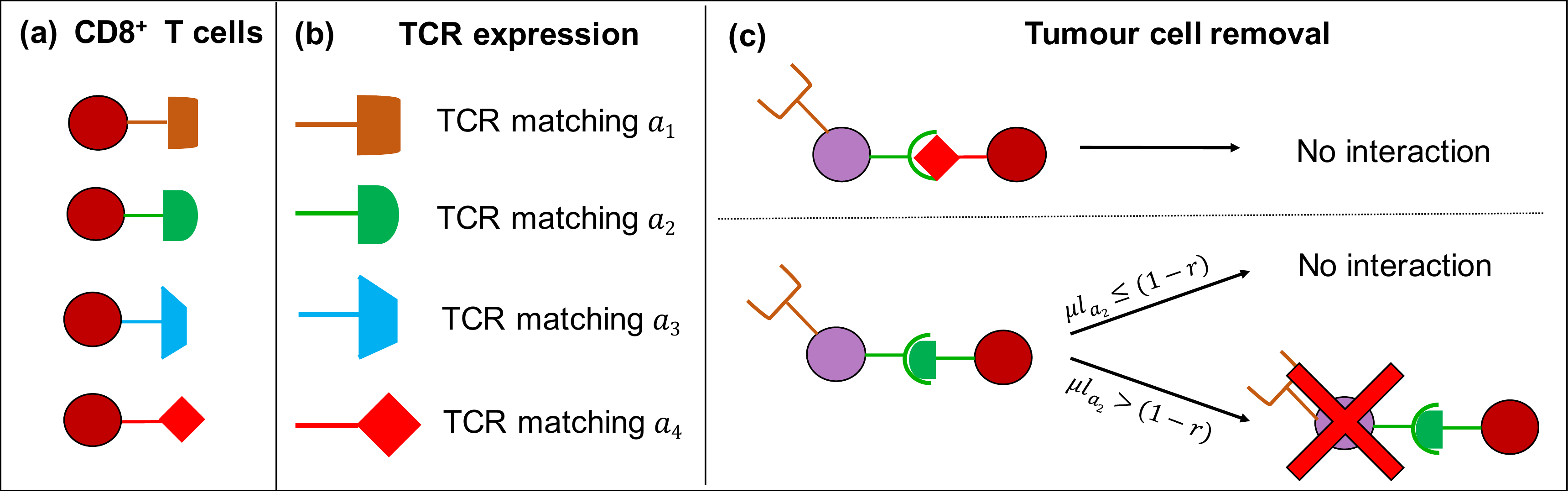}
    \caption{\textbf{Schematic representation of the modelling assumptions for CD8$^+$ T cells and their interaction with tumour cells.} \textbf{(a)} Red circles represent CD8$^+$ T cells, which express a unique TCR. \textbf{(b)} TCR are represented with different shapes and colors. Each TCR is able to recognize a particular tumour antigen. In the model, the number of TCRs is equal to the number of expressed tumour antigens. \textbf{(c)} Purple circles represent tumour cells. CD8$^+$ T cells can eliminate tumour cells, upon contact, under certain conditions. A tumour cell is eliminated if it presents the antigen matching the CD8$^+$ T cell receptor at a sufficiently high level. In this example, a tumour cell expressing antigen $a_1$ and $a_2$ cannot be eliminated by a CD8$^+$ T cell with TCR matching antigen $a_4$. On the other hand, the same tumour cell may be eliminated by a CD8$^+$ T cell expressing the TCR matching antigen $a_2$, under a condition on the level $l_{a_{2}}$ of presentation of such antigen $a_2$. The parameter $r$ is the intrinsic TCR-recognition probability and $\mu$ is a random variable drawn from a standard uniform distribution.}
    \label{fig:resumeHp_2}
\end{figure}

\paragraph{Influx of CD8$^+$ T cells}
 Following \citet{gong2017computational}, to model the tumour vessels that allow the arrival of CD8$^+$ T cells in the tumour micro-environment, we generate a set of points in the domain. In order not to rely on a detailed angioarchitecture, we generate $5$ entry points, equidistant from each other and from the centre of the domain. At each time step, a CD8$^+$ T cell specific to antigen $a_i\in A$ can be supplied to the domain from one of the 5 entry points, provided that the entry point is not occupied by other cells. The probability $0<p(t)\leq 1$ of influx of a CD8$^+$ T cell specific to antigen $a_i$ into the domain is proportional to the total amount $S_{a_i}^{tot}(t)$ of chemoattractant associated to antigen $a_i$ secreted at time $t$. Therefore, we define $p(t)$ as
 $$p(t):=C_2\,S_{a_i}^{tot}(t),$$ with $C_2 \in \mathbb{R^+}$ a scaling factor of units $\frac{[\textit{time}]}{[\textit{mol}]}$. \\
 Since the secretion of chemoattractants by tumour cells is proportional to the level of antigen presentation (see Eq.~\eqref{antigen_pres}), the total amount of chemoattractants secreted by non-immunogenic cells is lower than the total amount of chemoattractants secreted by immunogenic cells. Therefore, the influx of CD8$^+$ T cells targeted to sub-clonal antigens, which are expressed only by non-immunogenic cells, is lower than the influx of CD8$^+$ T cells targeted to clonal antigens.
\paragraph{TCR expression and T cell death}
We denote by $N_C(t)$ the number of CD8$^+$ T cells in the system at time $t$, and we label each of them by an index $m= 1,...,N_{C}(t)$.
Every CD8$^+$ T cell has a unique TCR [see Figure \ref{fig:resumeHp_2}(a)], and we suppose that each TCR is specific to a unique tumour antigen [see Figure \ref{fig:resumeHp_2}(b)]. When the $m^{th}$ CD8$^+$ T cell with a TCR targeted against antigen $a_i\in A$ arrives into the domain it undergoes chemotactic movement toward tumour cells expressing the matching antigen $a_i$. \\
CD8$^+$ T cell division occurs mostly in the lymph nodes \citep{delves2000immune} and cells then move to the tumour site. CD8$^+$ T cells can also proliferate at the tumour site but this is not the main site of proliferation. We thus neglect the effects of CD8$^+$ T cell proliferation at the tumour site and consider only the effects of proliferation outside the spatial domain of the model, leading to a varying influx of CD8$^+$ T cells.
A CD8$^+$ T cell undergoes apoptosis when it reaches the end of its intrinsic lifespan, which is drawn from a uniform distribution upon its arrival in the domain. 
\paragraph{Elimination of tumour cells by CD8$^+$ T cells}
Upon contact, CD8$^+$ T cells can interact only with tumour cells expressing the matching antigen [see Figure \ref{fig:resumeHp_2}(c)], and can induce their death, on the condition that the matching antigen is presented at a sufficiently high level. If a CD8$^+$ T cell is in contact with more than one tumour cell expressing the matching antigen, it will try to eliminate the one presenting the antigen at the highest level. In particular, when the $m^{th}$ CD8$^+$ T cell interacts with the $n^{th}$ tumour cell expressing the matching antigen $a_i$, we let the tumour cell be removed from the system, provided that
\begin{equation}\label{eq:killingproba}
\mu \, l_{a_i}^n>(1-r).
\end{equation}
Here $\mu$ is a random variable drawn from the standard uniform distribution, $l_{a_i}^n$ is the level of presentation of antigen $a_i$ by the $n^{th}$ tumour cell and $0<r\leq 1$ is the intrinsic TCR-recognition probability, which we suppose to be equal for every CD8$^+$ T cell. If the tumour cell satisfies the conditions to be eliminated, it undergoes apoptosis. The parameter $r$ determines the range of tumour cells the CD8$^+$ T cell population can interact with: large values of $r$ represent a CD8$^+$ T cell population able to eliminate tumour cells presenting their antigens at a low level, whereas low values of $r$ model the scenario where the CD8$^+$ T cells can only eliminate tumour cells presenting their antigens at a high level. \\
Tumour cell elimination by CD8$^+$ T cells takes approximately 6 hours to be completed \textit{in vitro} \citep{caramalho2009visualizing} and \textit{in vivo} \citep{breart2008two}. Accordingly, we require that an elimination event keeps a CD8$^+$ T cell engaged for 6 hours and only after this time the CD8$^+$ T cell can eliminate again \citep{kather2017silico}. If the condition \eqref{eq:killingproba} is not satisfied, the CD8$^+$ T cell is not engaged and can try, in the next time step, to eliminate again a tumour cell.

\subsubsection{Chemoattractant field}
\label{chemoattractant field}
As mentioned earlier, we let the $n^{th}$ tumour cell at the border of the tumour secrete a different chemoattractant for each antigen $a_i$ that it expresses. 
Denoting by $c_{a_i}$ the concentration of the chemoattractant secreted by tumour cells expressing antigen $a_i$, we let the dynamic of $c_{a_i}$ be described by the following reaction-diffusion equation:
\begin{equation}
\frac{\partial c_{a_i}}{\partial t}= D \Delta c_{a_i}- \gamma c_{a_i}+ \sum _{n\in N_{BT}(t)}s_{a_i}^n, \quad a_i\in A.
\label{chemoattr}
\end{equation}
In Eq.~\eqref{chemoattr}, $D$ is the diffusion constant and $\gamma$ is the rate of natural decay; these two parameters are assumed to have the same value for each chemoattractant. On the other hand, we recall that the secretion rate $s_{a_i}^n$ is specific to the $n^{th}$ tumour cell, because it is proportional to the level of presentation of antigen $a_i$ by the tumour cell (see Eq.~\eqref{antigen_pres}). $N_{BT}(t)$ denotes the set of tumour cells in contact with the surrounding medium at time $t$. \\
We add to Eq. \eqref{chemoattr} zero-flux boundary conditions and an initial concentration $c_a^{init}$ which is set to be zero everywhere in the domain but at the border of the tumour.

\section{Numerical simulations and preliminary results}
\label{Computational simulations}
\subsection{Set-up of simulations}
For numerical simulations of our individual-based model, we use a Cellular Potts approach on a 2D spatial grid with a total of $400\times 400$ lattice sites. Simulations were developed and run using the software CompuCell3D \citep{izaguirre2004compucell} on a standard workstation (Intel i7 Processor, 4 cores, 16 GB RAM, macOS 11.2.2), with one time-step chosen to be $\Delta t = 1 $ min. The computational implementation of Cellular Potts models is described in \ref{Details of computational model}, while full details of the model parametrisation are provided in \ref{appendix:graph}. Files to run a simulation example with Compucell3D software \citep{izaguirre2004compucell} are available at: \url{https://plmlab.math.cnrs.fr/audebert/cc3dmodeltumourcd8}.  
 \\
At the initial time point of the simulation, a certain number of tumour cells are already present in the domain, while CD8$^+$ T cells arrive only when the simulation starts. At the beginning of simulations there is a total of $400$ tumour cells, tightly packed in a circular configuration positioned at the centre of the domain, reproducing the geometry of a solid tumour.\\
All quantities we present in this section and in Section \ref{Main results} are obtained by averaging over the results of $10$ simulations, with parameter values kept constant and equal to those listed in Table \ref{table1} and Table \ref{table2}. Unless otherwise explicitly stated, we carry out numerical simulations for 28800 time-steps, corresponding to 20 days. \\
The next two subsections describe two preliminary computational results of our model which will be used to guide the simulations leading to the main results presented in Section \ref{Main results}. 
\subsection{Baseline scenario: tumour development in the absence of CD8$^+$ T cells}
\label{baseline scenario}
\begin{figure}[!t]
\centering
\includegraphics[width=12cm]{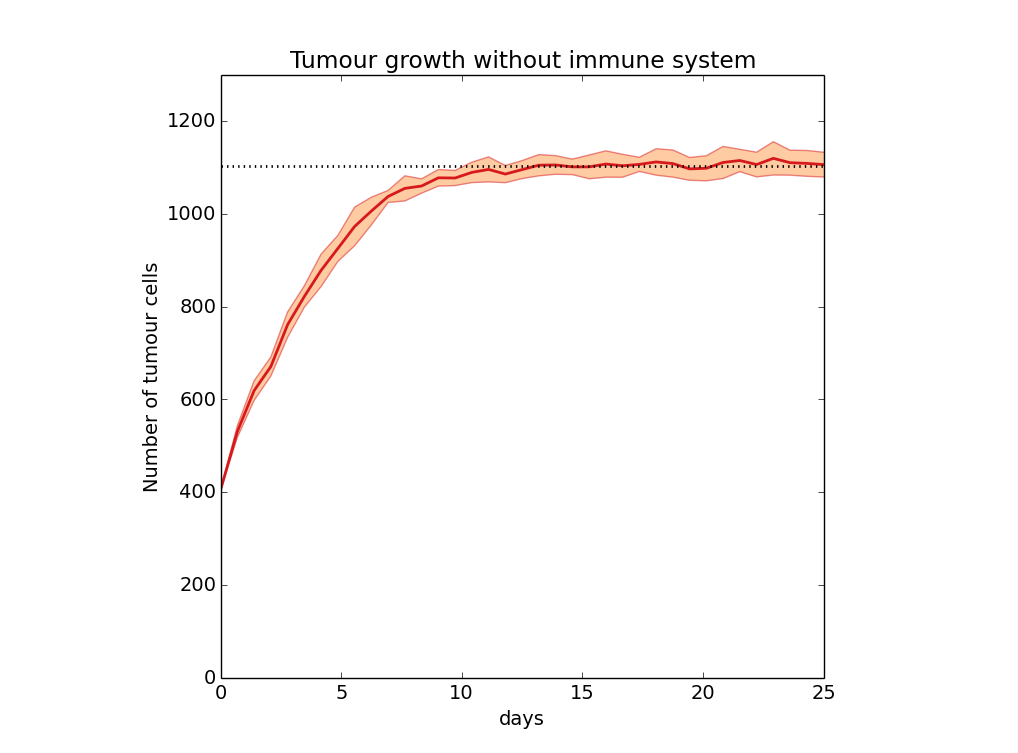}
\caption{\textbf{Baseline scenario: tumour development in the absence of CD8$^+$ T cells.} Time evolution of the tumour cell number in the absence of CD8$^+$ T cells. The shaded area indicates $+/-$ standard deviation between 10 simulations. The black dotted line highlights a numerical estimation of the tumour cell carrying capacity.}
\label{tumour_no_T_cells}
\end{figure}

We first establish a baseline scenario where tumour cells grow, divide and die via the modelling rules described in Section \ref{Tumour cells}, in the absence of CD8$^+$ T cells. For this case, we carry out numerical simulations for $36000$ time-steps, corresponding to $25$ days. Figure \ref{tumour_no_T_cells} shows the growth over time of the number of tumour cells. The growth of the tumour cell number is of logistic type, as expected by the rules that govern tumour cell death. Logistic growth has been used by a number of authors to model the temporal evolution of the size of solid tumours \citep{drasdo2003individual,kirschner1998modeling,kuznetsov1994nonlinear}. 
The carrying capacity, \textit{i.e.} the saturation value reached by the number of tumour cells due to intra-population competition, is numerically estimated to be of about $1100$ cells.\\
In the following subsections, we explore the immune response to tumours characterised by different degrees of ITH. Each simulation is carried out by keeping all parameter values fixed (and equal to this baseline scenario) and changing only the initial compositions of the different tumours.
\subsection{Tumours with larger number of sub-populations of cancer cells lead to lower immune response efficacy}
\label{Larger number of sub-populations corresponds to a decreased efficacy of the immune response}
In a recent \textit{in vivo} study in mice, the volume of UVB irradiated tumours after 20 days is found to be linked to the number of sub-populations of cancer cells constituting each inoculated tumour \citep{wolf2019uvb}. In our next simulations, we attempt to verify that our model reproduces such phenomenon,  exploring the outcomes of immune response to 7 different tumours characterised by an increasing number of sub-populations of tumour cells. At this preliminary stage, we simplify our model. We consider that each tumour consists of the same type of cells, and we do not differentiate between immunogenic and non-immunogenic cells. In particular, here we let each sub-population of tumour cells be characterised by cells expressing a single antigen. This antigen is presented at a random level, chosen uniformly from the discrete set $L=\left\{\frac{1}{6},\frac{2}{6},\frac{3}{6},\frac{4}{6},\frac{5}{6},1\right\}$. In this way, tumour cells that express the same antigen belong to the same sub-population. \\

\begin{figure}[!t]
\centering
\includegraphics[width=17cm]{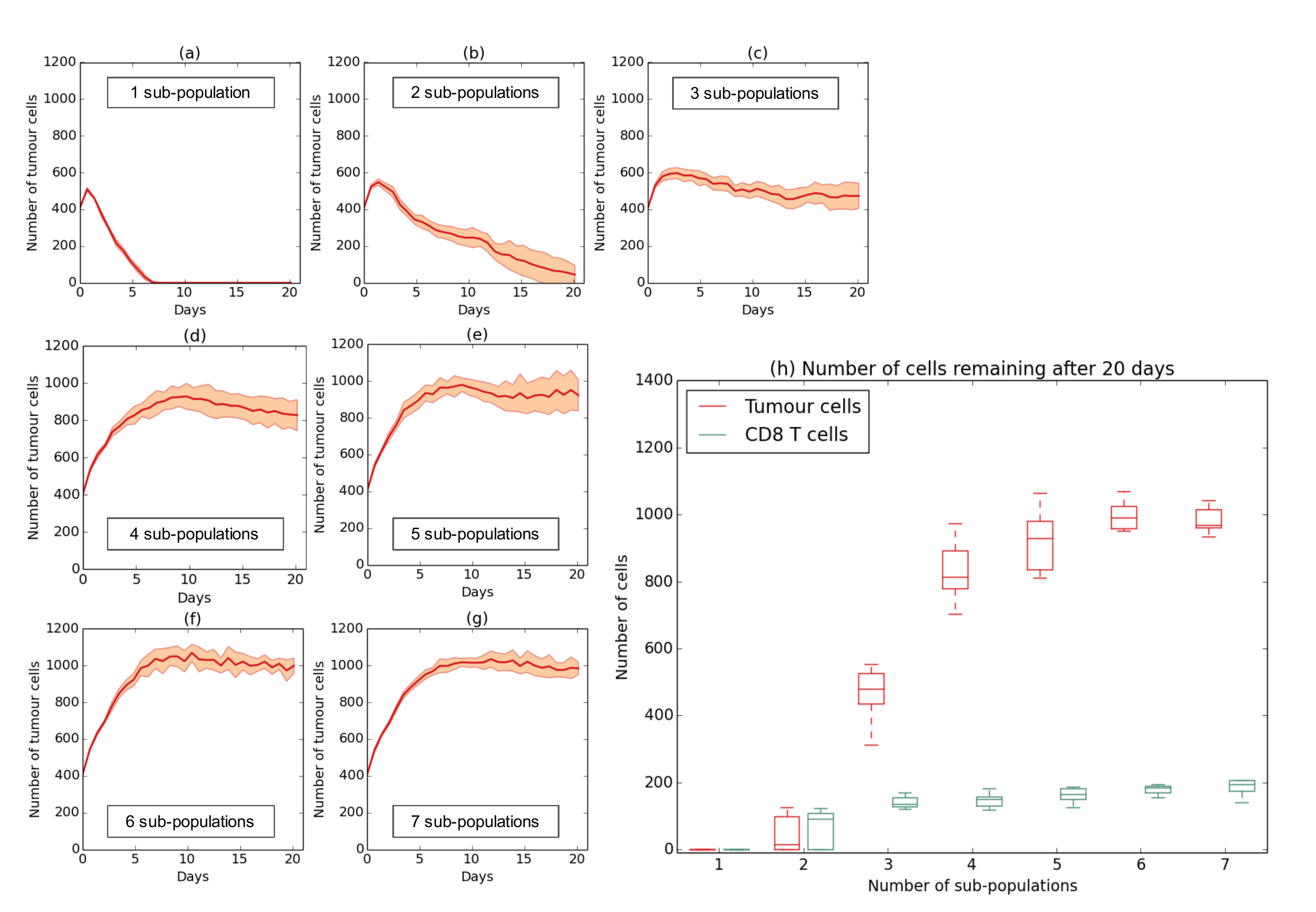}
\caption{\textbf{Tumours with larger number of sub-populations of cancer cells lead to lower immune response efficacy.} Plots in panel \textbf{(a)-(g)} display the time evolution of the tumour cell number for tumours characterised by increasing numbers of sub-populations. Shaded areas indicate $+/ -$ standard deviation between 10 simulations. Plot in panel \textbf{(h)} displays the corresponding number of tumour cells (in red) and CD8$^+$ T cells (in green) remaining after 20 days (28800 time-steps) for the different initial tumour compositions. The cell numbers presented here were obtained as the average over 10 simulations and the error bars display the related standard deviation.}
\label{differentPopulations}
\end{figure}

The plots in Figure \ref{differentPopulations}(a)-(g) display the time evolution of the tumour cell number of 7 tumours that comprise 1 to 7 different sub-populations of cancer cells. Plot in Figure \ref{differentPopulations}(h) displays the corresponding number of tumour cells and CD8$^+$ T cells remaining at the end of simulations (after 20 days) for the 7 tumours. For tumours constituted of 1 or 2 sub-populations of cancer cells, none or very few tumour cells remain after 20 days [Figure \ref{differentPopulations}(a)-(b)]. When 3 sub-populations of cancer cells constitute the tumour, the number of tumour cells over time tends to stay constant and slightly above its initial value [Figure \ref{differentPopulations}(c)]. Finally, for tumours initially constituted of more than 3 sub-populations, the number of tumour cells after $20$ days is more than twice the initial value. In addition, the final number of tumour cells increases as we increase the number of sub-populations of cancer cells constituting the tumour from 1 to 6. For tumours with 6 and 7 sub-populations, the final number of tumour cells is similar and is about $1000$ cells [Figure \ref{differentPopulations}(f)-(g)]. 
These results support the idea that the anti-tumour immune action is efficient only when the tumour is constituted of 1 or 2 sub-populations of tumour cells. Moreover, up to a certain point, the increase of the number of sub-populations of tumour cells results in a weaker immune response. Finally, increasing the number of sub-populations of tumour cells beyond 6 does not appear to change the effect of the immune response. Comparing the dynamics of the two last tumours (with 6 and 7 sub-populations of tumour cells) to the baseline scenario of Section \ref{baseline scenario}, we see that the immune response is almost inefficient, as it is not able to really limit the growth of the two tumours.\\
Our computational results are in agreement with experimental results presented in \citep{wolf2019uvb}. In this study, the induction of UVB-derived tumour, which lead to an increase in the number of sub-populations of cancer cells, results in aggressive tumours with reduced anti-tumour immune activity. However, when different single-cell-clone derived tumours (characterised by a unique sub-population of tumour cells) are considered, the immune system is able to effectively eradicate them.\\

Figure \ref{differentPopulations}(h) also shows that, when the tumour is constituted of more than 3 sub-populations of tumour cells, the total number of CD8$^+$ T cells at the end of simulations remain almost constant in the different tumours (around $200$ cells). A consequence of this is that the average size of each specific CD8$^+$ T cell sub-population decreases as we consider tumours with increased number of sub-populations of tumour cells. This leads to a less efficient anti-tumour immune response.  As highlighted by \citet{wolf2019uvb}, these computational results also suggest that increasing the number of sub-populations of tumour cells reduce the exposition of each antigen to the ``front-line'', thus making more difficult for immune cells to detect them. The resulting outcome is a reduced influx of specific CD8$^+$ T cells in the tumour micro-environment and a weaker anti-tumour immune response.

\subsection{Initial composition of two tumours inspired by biological studies}
\label{Initial conditions}
In the previous subsections we investigated simple cases of tumour growth with and without the action of the immune system. We will now explore further the effect of ITH on immune surveillance considering two tumours inspired by biological studies, in order to effectively capture more layers of biological complexity. For the two tumours we consider different initial antigenic compositions, corresponding to different degrees of ITH. In particular, following the experiments presented in \citep{wolf2019uvb}, we dissect out two characteristics of ITH: the number of sub-populations of cancer cells constituting a tumour and the percentage of immunogenic and non-immunogenic cells within it. With our model, we wish to investigate the effect of these two expressions of ITH on tumour aggressiveness independently and together, evaluating their influence on anti-tumour immunity in a controlled manner. 
To this end, first we generate two tumours with different number of sub-populations of cancer cells. Following the experiments presented in \citep{wolf2019uvb}  and the results of Section\ref{Larger number of sub-populations corresponds to a decreased efficacy of the immune response}, we consider, respectively, tumours with 3 and 7 sub-populations of cancer cells. For simplicity, we denote the first tumour as \textit{tumour-3a} and the second one as \textit{tumour-7a}. The antigenic composition of the two tumours and their corresponding phylogenetic tree representation are inspired by \citet{wolf2019uvb}. More details about the two tumours are given in the next paragraphs. Next, for each tumour we consider different initial percentages  of immunogenic and non-immunogenic cells. When different sub-populations of immunogenic (or non-immunogenic) cells are considered, the total percentage of immunogenic (or non-immunogenic) cells is equally distributed in each sub-population. This enables us to decouple antigen heterogeneity and antigen immunogenicity, and study their influence on tumour aggressiveness in a causal, systematic manner. 
\begin{figure}[!t]
    \centering
    \includegraphics[scale=0.55]{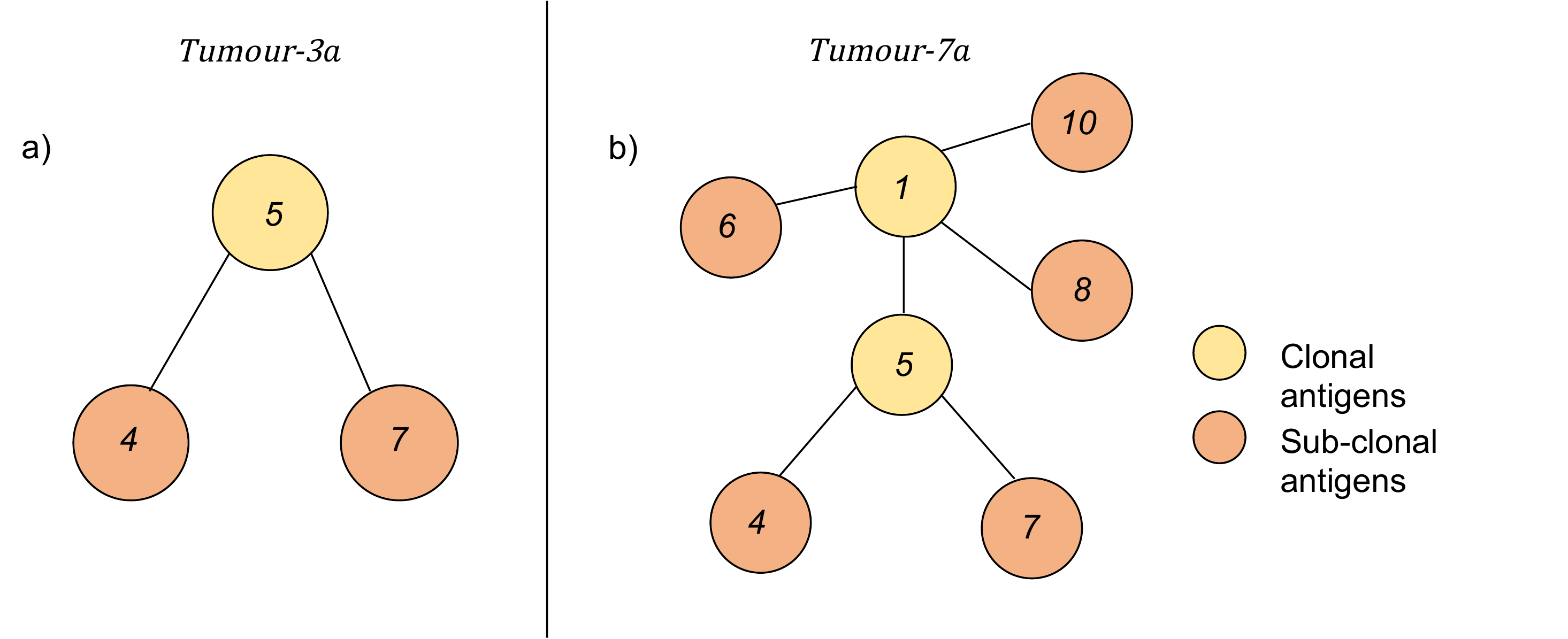}
    \caption{\textbf{Phylogenetic tree representations of the antigens considered for the two tumours.} \textbf{(a)} \textit{Tumour-3a} expresses three antigens. Antigen 5 is the only clonal antigen (in yellow) and antigens 4 and 7 are two sub-clonal antigens (in orange). As a results, \textit{tumour-3a} is composed of 3 sub-populations of tumour cells. \textbf{(b)} \textit{Tumour-7a} expresses seven antigens. Antigens 1 and 5 are clonal antigens (in yellow) and antigens 4, 6, 7, 8 and 10 are sub-clonal antigens (in orange). Hence, \textit{tumour-7a} is composed of 7 sub-populations of tumour cells. The phylogenetic tree representations of the two tumours are inspired by \citet{wolf2019uvb}.}
    \label{fig:homHeter}
\end{figure}
\paragraph{Tumour-3a}
\label{Homogeneous and clonal tumour}
The first tumour we consider expresses three different antigens, one of which is clonal and the other two are sub-clonal (see Figure \ref{fig:homHeter}(a)). With the notation introduced in Section \ref{Tumour cells}, we denote respectively by
\begin{equation}
    A=\{4,5,7\}, \quad A_C=\{5\} \quad \text{and} \quad A_{SC}=\{4,7\}
\end{equation}
the antigen profile of the tumour, the clonal antigens and the sub-clonal antigens. \\
 Based on the phylogenetic tree representation of Figure \ref{fig:homHeter}(a), we divide \textit{tumour-3a} in 3 sub-populations of tumour cells labelled by the last antigen acquired by each cell. Cells in the sub-population labelled by antigen $5$ carry only this antigen, while cells in sub-populations labelled by antigens $4$ and $7$ express, respectively, antigens $5$ and $4$ or antigens $5$ and $7$. \\
As cells in the sub-population labelled by antigen $5$ are immunogenic, their level of antigen presentation by the MHC-I is assumed to be normal and randomly chosen from the discrete set $L_{I}$, which is defined as
\begin{equation}
    L_{I}=\left\{\frac{1}{6},\frac{2}{6},\frac{3}{6},\frac{4}{6},\frac{5}{6},1\right\}. \label{Lc}
\end{equation} 
On the other hand, as cells in sub-populations labelled by antigens $4$ and $7$ are non-immunogenic, their level of antigen presentation by the MHC-I is deteriorated and, therefore, randomly chosen from the discrete set $L_{NI}$, which is defined as \begin{equation}
  L_{NI}=\left\{\frac{5}{100},\frac{10}{100},\frac{15}{100},\frac{20}{100},\frac{25}{100},\frac{30}{100}\right\}. 
  \label{Lsc}
\end{equation}  
\paragraph{Tumour-7a}
\label{Heterogeneous and sub-clonal tumour}
The second tumour expresses seven different antigens, two of which are clonal and five are sub-clonal (see Figure \ref{fig:homHeter}(b)). We denote respectively by
\begin{equation}
    A=\{1,4,5,6,7,8,10\}, \quad A_C=\{1,5\} \quad \text{and} \quad A_{SC}=\{4,6,7,8,10\}
\end{equation}
the antigen profile of the tumour, the clonal antigens and the sub-clonal antigens. \\
Based on the phylogenetic tree representation of Figure \ref{fig:homHeter}(b), we divide \textit{tumour-7a} in 7 sub-populations of tumour cells. 
Cells in sub-populations labelled by antigens $1$ and $5$ are immunogenic, and present their antigens at a level randomly chosen from set $L_{I}$, which is defined in \eqref{Lc}. On the other hand, cells in sub-populations labelled by antigens $4$, $6$, $7$, $8$ and ${10}$ are non-immunogenic, and present all their antigens at a lower level randomly chosen from set $L_{NI}$, which is defined in \eqref{Lsc}.\\

In the next Section, we investigate the effects of CD8$^+$ T cell response to different tumours characterised by different levels of ITH. The obtained dynamics are compared with the baseline scenario.  In the next simulations, we consider different compositions of the initial tumour, while the other parameters are kept constant to the values listed in Table \ref{table1} and Table \ref{table2}. The values of the parameters are chosen so as to qualitatively reproduce essential aspects of the experimental results obtained by \citet{wolf2019uvb}
\section{Results and discussion}
\label{Main results}
\subsection{Large number of sub-populations of cancer cells constituting a tumour reduces the effectiveness of the immune response}
\label{The number of sub-populations of cancer cells constituting  a tumour correlates with the effectiveness of the immune response}
To investigate how the immune response is affected by different degrees of heterogeneity, we start by comparing two situations in which the initial tumours are characterised by different number of sub-populations of tumour cells. We consider as initial conditions \textit{tumour-3a}, with 3 different sub-populations of tumour cells, and \textit{tumour-7a}, with 7 different sub-populations of tumour cells, defined as in Section \ref{Initial conditions}. For each tumour, we consider the same initial percentage of immunogenic and non-immunogenic cells, corresponding to $75\%$ of immunogenic cells and $25\%$ of non-immunogenic cells. Note that these two tumours are different from those considered in the results presented in Section \ref{Larger number of sub-populations corresponds to a decreased efficacy of the immune response}. In fact, in \textit{tumour-3a} and \textit{tumour-7a} cells can express clonal and sub-clonal antigens and, therefore, be either of immunogenic or non-immunogenic type. On the other hand, in the results presented in Section \ref{Larger number of sub-populations corresponds to a decreased efficacy of the immune response}, each tumour cell presents a unique antigen, which is shared by all the cells in the same sub-population of tumour cells. The situation considered here provides a more faithful representation of biological complexity, as a tumour cell can express more than one antigen presented at different levels.\\

\begin{figure}[!t]
\includegraphics[width=16.7cm]{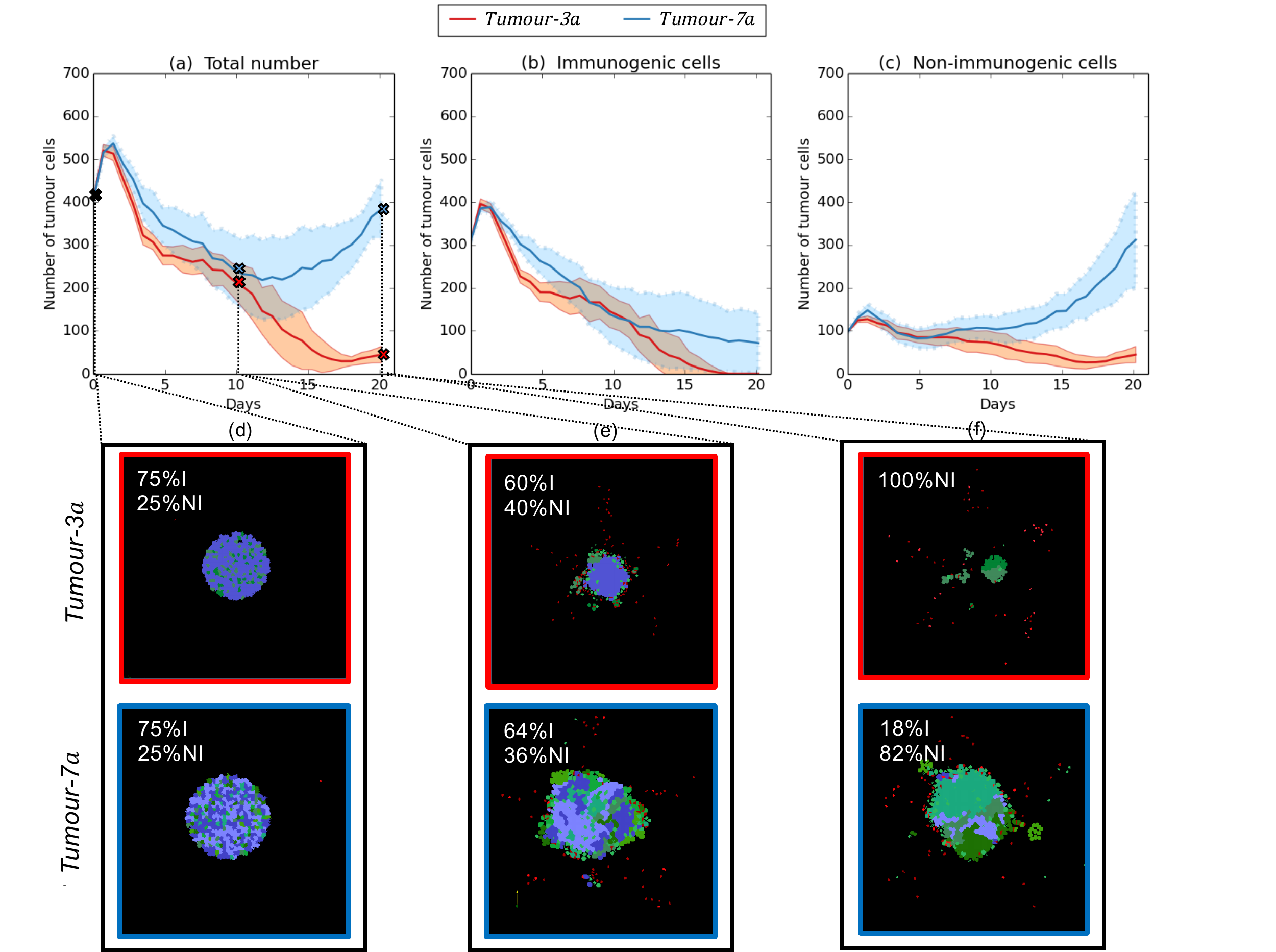}
\caption{\textbf{The number of sub-populations constituting a tumour impacts on the effectiveness of the immune response.} Plots in panels \textbf{(a)-(c)} display the time evolution of the total tumour cell number, and the corresponding evolution of the number of immunogenic cells and non-immunogenic cells for \textit{tumour-3a} (in red) and \textit{tumour-7a} (in blue). Shaded areas indicate $+/-$ standard deviation between 10 simulations. For these simulations, an equal initial percentage of $75\%$ of immunogenic cells and $25\%$ of non-immunogenic cells was considered. Insets in panels \textbf{(d)-(f)} display an example of the spatial distribution of cells for \textit{tumour-3a} (first row) and \textit{tumour-7a} (second row) at different times of the simulation. Purple cells are immunogenic cells, green cells are non-immunogenic cells and red cells are CD8$^+$ T cells.}
\label{population}
\end{figure}

Figure \ref{population}(a)-(c) show the time evolution of the total number of tumour cells, along with the corresponding time evolution of immunogenic and non-immunogenic cell number. Figures \ref{population}(d)-(f) also display the spatial cell distributions observed at different times of two simulations.
As shown by Figure \ref{population}(a), the two tumours have similar dynamics from the beginning of simulations until day $10$, with an initial increase of the cell number followed by a steep decrease. After day 10, in \textit{tumour-3a}, the number of tumour cells continues to decrease until it reaches a low, almost constant level. 
Figure \ref{population}(b) and (c), along with the corresponding panel of Figure \ref{population}(f), show that, at the end of simulations, all the immunogenic cells are eliminated by the CD8$^+$ T cells, and only few non-immunogenic cells remain in the system.
On the other hand, for \textit{tumour-7a}, after day $10$ the tumour cell number increases steadily over time. This dynamic leads to a final tumour size similar to the initial one. Moreover, as shown by Figure \ref{population}(b), the number of immunogenic cells decreases over time, whereas the number of non-immunogenic cells, after being initially kept under control by immune cells, increases steeply (Figure \ref{population}(c)). The related panels of Figure \ref{population}(d)-(f) show the progressive colonisation of the tumour by non-immunogenic cells. \\
Comparing these results with the baseline scenario of Section \ref{baseline scenario}, for both tumours we clearly see the effects of the action of immune cells on tumour growth, which is no longer simply logistic and saturating to carrying capacity. However, the effectiveness of the immune response depends on the tumour considered. For \textit{tumour-3a}, the immune response is efficient and almost eliminates all tumour cells. On the other hand, for \textit{tumour-7a}, the higher heterogeneity leads to a less effective immune response and the tumour eventually grows again. These results suggest that, even if characterised by equal percentages of immunogenic and non-immunogenic cells, tumours with a larger number of sub-populations of tumour cells, which express a wider spectrum of antigens, are more aggressive. This was already suggested by the results presented in Section  \ref{Larger number of sub-populations corresponds to a decreased efficacy of the immune response}. Moreover, even with all tumour cells presenting clonal antigens (it was not the case in Section \ref{Larger number of sub-populations corresponds to a decreased efficacy of the immune response}) the CD8$^+$ T cells are not able to control the growth of the tumour. This again indicates that the number of sub-populations and antigens in a tumour have an impact on the effectiveness of the immune response.
\\

The outcomes of our model indicate that in \textit{tumour-3a} the presence of a low number of antigens leads to a better immune detection, enhancing the ability of the immune system to eliminate the tumour. In both tumours the immune system rapidly targets and eliminates immunogenic cells, giving a competitive advantage to non-immunogenic cells. In fact, we initially observe a reduction in the number of tumour cells.
However, in \textit{tumour-7a}, as more sub-populations of tumour cells are present, non-immunogenic cells have a better chance of escaping immune surveillance. The outcome is a weaker anti-tumour immune response.
Overall, our results are in agreement with the recent hypothesis by \citet{wolf2019uvb} that, because of increased antigenic variability, the relative expression of each antigen is weaker in tumours composed of a larger number of sub-populations of tumour cells. In particular, clonal antigens undergo ``dilution''  within the tumour, and, therefore, the chance for CD8$^+$ T cells to identify immunogenic cells is reduced. This leads to a diminished ability of CD8$^+$ T cells to mount a sufficient cytotoxic response. 

\subsection{Different initial percentages  of immunogenic and non-immunogenic cells can cause variations in anti-tumour immune response} 

The results discussed in the previous subsection illustrate how the effectiveness of the immune response can decreases in tumours with larger number of sub-populations of tumour cells. We investigate the effects of ITH further, focusing on the role of the percentage of immunogenic and non-immunogenic cells. We fix the number of sub-populations of tumour cells considering only \textit{tumour-3a}, and vary the initial percentage of immunogenic and non-immunogenic cells.\\

\begin{figure}[!t]
\includegraphics[width=15cm]{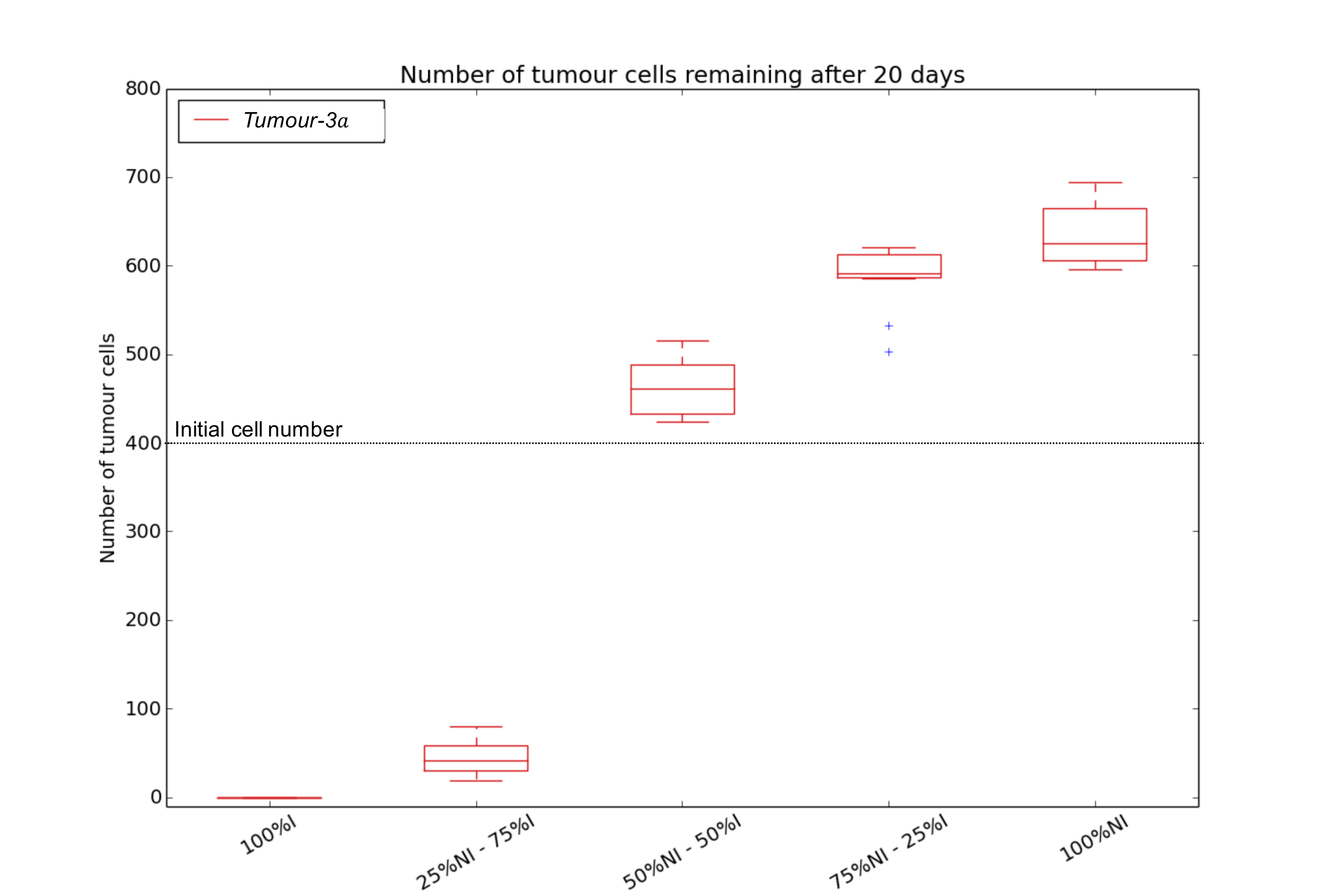}
\centering
\caption{\textbf{Different initial percentages  of immunogenic and non-immunogenic cells can cause variations in the immune response to tumour cells.} Plot displaying the number of tumour cells remaining after 20 days (28800 time-steps) for different initial percentages  of immunogenic and non-immunogenic  cells. For these simulations, only \textit{tumour-3a} was considered. The tumour cell numbers presented here were obtained as the average over 10 simulations and the error bars display the related standard deviation. The black dotted line highlights the number of tumour cell at the initial time of the simulations.}
\label{boxplot_homogeneous}
\end{figure}

The plot in Figure \ref{boxplot_homogeneous} displays the number of tumour cells remaining at the end of simulations (after 20 days), for different initial percentages  of immunogenic and non-immunogenic cells. For low percentages  of non-immunogenic cells ($\leq 25\%$), none or very few tumour cells survive after 20 days. On the contrary, for tumours initially composed of more than $50\%$ of non-immunogenic cells, the number of tumour cells after $20$ days is larger than the initial one. 
In addition, the final number of cells increases as we increase the initial percentage of non-immunogenic cells. These results suggest that the anti-tumour immune action is efficient only when the percentage of non-immunogenic cells is low compared to the percentage of immunogenic cells. Moreover, the larger the percentage of non-immunogenic cells, the weaker the immune response is.\\

Compared to the baseline scenario of Section \ref{baseline scenario}, we see the effects of the immune system on tumour growth. In fact, for each scenario the number of cells at the end of simulations is lower than the tumour carrying capacity shown in Figure \ref{tumour_no_T_cells}. However, for larger percentages  of non-immunogenic cells, the immune response is not efficient enough to reduce the initial tumour size. \\

Taken together, our results qualitatively reproduce key findings of experiments performed in \textit{in vivo} syngeneic mice tumour models \citep{gejman2018rejection}. The results presented in \citep{gejman2018rejection} indicate that a non-effective immune response may occur when the percentage of immunogenic cells in the tumour is low. Our computational results provide an explanation for such emergent behaviour. Since sub-clonal antigens are presented at a low level by the MHC-I, non-immunogenic cells trigger a poor CD8$^+$ T cell response. Thus, tumours characterised by a major percentage of non-immunogenic cells result in a weaker overall immune response.
Furthermore, \citet{gejman2018rejection} put forward the idea that the threshold percentage of immunogenic cells that is required to trigger an antigen-specific CD8$^+$ T cell response may vary depending on the antigens. In order to address this point, such a feature could be implemented in the model, for example by considering antigen presentation levels or chemotactic responses specific to each antigen.\\

\begin{figure}[!t]
\includegraphics[width=16cm]{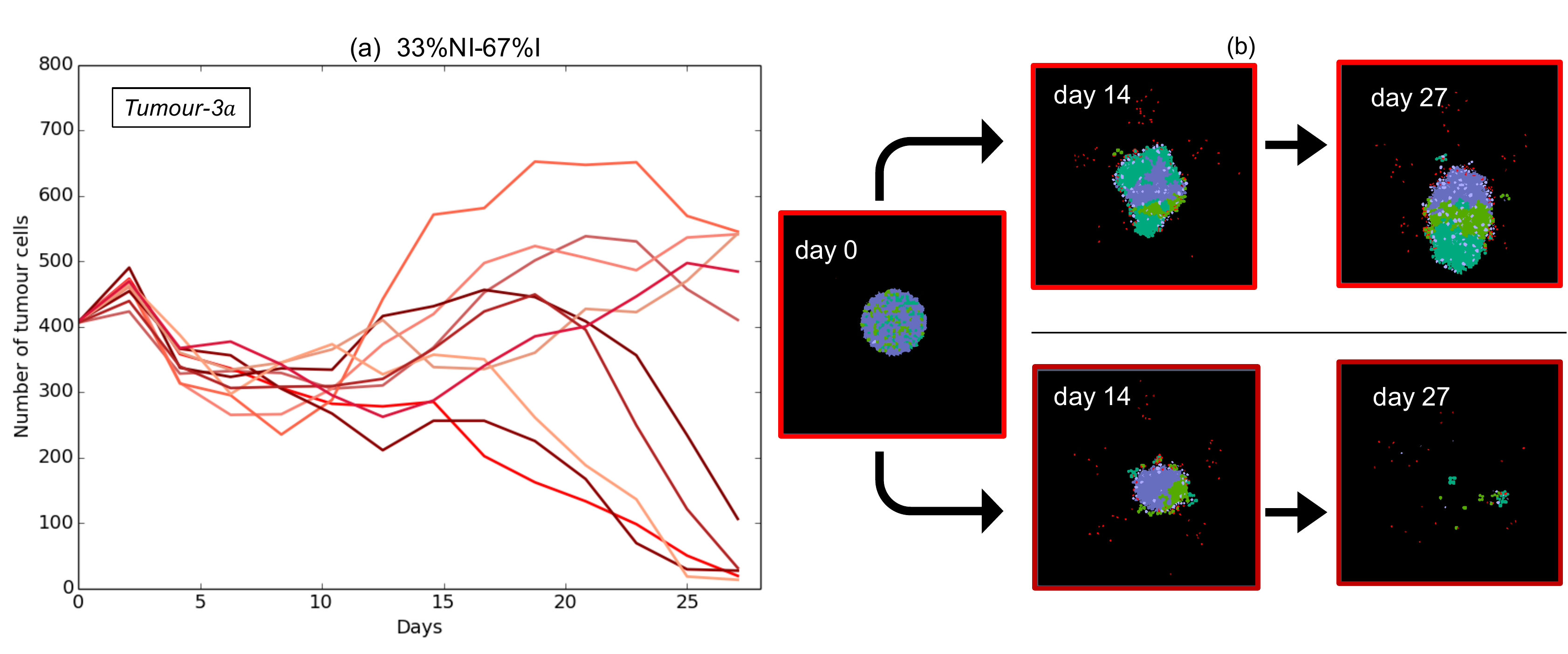}
\centering
\caption{\textbf{Stochasticity in cell dynamics may affect the outcomes of immune action.} Plot in panel \textbf{(a)} displays the time evolution of the tumour cell number for an initial percentage of $33\%$ of non-immunogenic cells and $67\%$ of immunogenic cells for 10 runs of simulations. Each line corresponds to a unique realisation of our model. For these simulations, only \textit{tumour-3a} was considered. The insets in panel \textbf{(b)} show an example of the observed spatial distributions of cells corresponding to different times of two simulations. }
\label{33_stoch}
\end{figure}

The role of the immunogenic cell percentage within the tumour is further analysed as we observe a gap between the results obtained considering $25\%$ and $50\%$ of non-immunogenic cells (see Figure \ref{boxplot_homogeneous}). This is investigated by performing simulations considering percentages  of non-immunogenic cells between these two values. Figure \ref{33_stoch} displays the time evolution of the number of tumour cells for 10 different realisations of the same simulation, considering the same initial condition with 33\% of non-immunogenic cells and 67\% of immunogenic cells. In this case, we carried out numerical simulations for 38800 time-steps (corresponding to 27 days). \\
Under this choice of the initial condition, we observe a large variability in the tumour-immune cell dynamics, which does not lead to a clear emergent behaviour. In particular, Figure \ref{33_stoch}(a) shows that, in some simulations, the number of tumour cells decreases over time and only few cells remain at the end of the simulations. In other cases, after an initial phase between day $0$ and day $10$ where CD8$^+$ T cells keep under control the growth of the tumour, the number of tumour cells eventually increases and the resulting final number of tumour cells is larger than the initial one. This is also illustrated by Figure \ref{33_stoch}(b), which displays a sample of the spatial cell distributions at different time of two simulations. In particular, here we show that, starting from the same initial condition, we obtain two different outcomes: in one case immune clearance occurs and tumour cells are almost entirely eliminated by the immune system; in the other case tumour cells escape immune surveillance. When immune escape occurs, in the example proposed in Figure \ref{33_stoch}(b) at day $14$, immunogenic cells are surrounded by non-immunogenic cells, which hamper immune detection. This leads to a decreased influx of CD8$^+$ T cells in the tumour micro-environment and results in a weaker immune response.\\
These results suggest that the stochasticity which is present in cell dynamics may affect the outcomes of immune action. These results may partially explain the outcomes of earlier experimental research \citep{chen2017elements, kim2012modeling}, which found that responses of patients with similar tumours can vary considerably. In this regard, the use of mathematical models for identification and understanding of immune escape mechanisms in individual tumour could help advancing personalized tumour treatment.

\subsection{Both the number of sub-populations of cancer cells constituting a tumour and the percentage of immunogenic and non-immunogenic cells affect the effectiveness of the immune response} 

So far, we have investigated with our model the effects of ITH on immune response by varying independently the number of sub-populations of cancer cells constituting a tumour and the percentage of immunogenic and non-immunogenic cells. Now, we study their combined effect in mediating tumour growth. We consider as initial conditions \textit{tumour-3a} and \textit{tumour-7a}, characterised by different numbers of sub-populations of tumour cells, and for different initial percentages  of immunogenic and non-immunogenic cells. \\

\begin{figure}[!t]
\includegraphics[width=16.5cm]{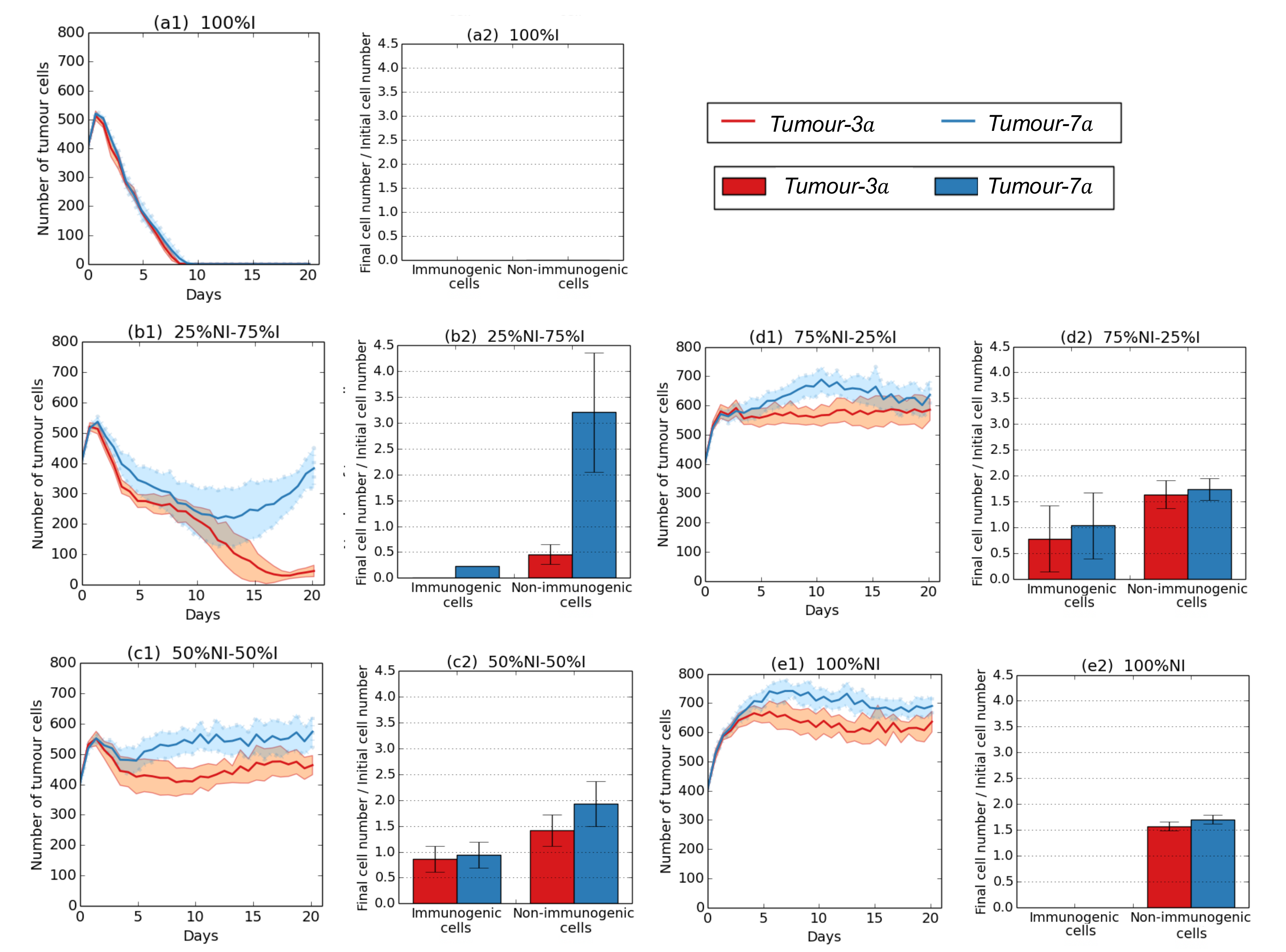}
\centering
\caption{\textbf{Both the number of sub-populations of cancer cells constituting  a tumour and the percentage of non-immunogenic cells affect the effectiveness of the immune response.} Plots in panel \textbf{(a1)-(e1)} display the time evolution of the tumour cell number for \textit{tumour-3a} (in red) and \textit{tumour-7a} (in blue). In both tumours, from \textbf{(a1)} to \textbf{(e1)} the initial percentage of non-immunogenic cells is increased. Shaded areas indicate $+/ -$ standard deviation between 10 simulations. Plots in panel \textbf{(a2)-(e2)} display the corresponding average number of immunogenic and non-immunogenic cells at the end of simulations with respect to the initial one. The error lines represent the standard deviation between $10$ simulations.}
\label{histo_plot}
\end{figure}

Figure \ref{histo_plot} displays the time evolution of the total number of cells for different initial tumour compositions, and compares the number of immunogenic and non-immunogenic cells at the end of simulations with respect to the initial one. As shown by Figure \ref{histo_plot}(a1), the immune system is able to completely eradicate the tumour only when it is initially composed of $100\%$ of immunogenic cells, independently of the number of sub-populations of tumour cells. When the initial tumour is made of $25\%$ of non-immunogenic cells, Figure \ref{histo_plot}(b1) show that the two tumours have different dynamics. In particular, as already observed in the results presented in Section \ref{The number of sub-populations of cancer cells constituting  a tumour correlates with the effectiveness of the immune response}, the number of cells in \textit{tumour-3a} decreases over time until the end of the simulations, while the number of cells in \textit{tumour-7a}, after an initial decrease, steadily increases until the end of the simulations. 
Finally, when tumours are initially composed of more than $50\%$ of non-immunogenic cells, similarly to the baseline scenario of Section \ref{baseline scenario}, they follow a logistic growth, except for an initial decrease shown by Figure \ref{histo_plot}(c1). For both tumours, the tumour cell number eventually saturates at a certain value (see Figure \ref{histo_plot}(c1)-(e1)). In these cases, the saturation value of the number of tumour cells is larger than the initial tumour cell number. Moreover, the saturation value attained increases as we increase the level of heterogeneity of the tumour (respectively, the number of sub-populations of tumour cells and the percentage of non-immunogenic cells). Such results indicate that in these cases CD8$^+$ T cells are present in the tumour micro-environment but do not produce an effective immune response. Persistent antigen presentation has been proven to cause continuous TCR stimulation that could directly induce CD8$^+$ T cell dysfunction and exhaustion \citep{yi2010t,zuniga2012t}. The model presented in this work does not include this aspect, but it could be easily extended to do so.\\

The outcomes of our model recapitulate the main results of in \textit{in vivo} clonal mixing experiments in mice models presented by \citet{wolf2019uvb}, who studied the combined effect of these two characteristics of ITH in mediating tumour growth and eradication. Wolf and collaborators have demonstrated that tumours with increased number of clones and large genetic diversity are more aggressive. In our model, the number of clones can be linked to the number of sub-populations of tumour cells, while genetic diversity may be linked to the immunogenicity of the tumour. Moreover, our findings are in agreement with an experimental work indicating that patients whose tumours are highly heterogeneous have increased levels of relapse after an initial response to immunotherapy and worse survival expectations than patients with more homogeneous tumours \citep{mcgranahan2016clonal}.\\
 
We next analyse the evolution over time of immunogenic and non-immunogenic cells. When tumours are initially composed of $25\%$ of non-immunogenic cells, Figure \ref{histo_plot}(b2) shows that the two tumours evolve in different ways. While the number of non-immunogenic cells is considerably reduced in \textit{tumour-3a}, the final number of non-immunogenic cells increases up to four times its initial number in \textit{tumour-7a}. On the other hand, when tumours are initially composed of more than $50\%$ of non-immunogenic cells, independently of the tumour considered, we observe a similar trend in the evolution of immunogenic and non-immunogenic cells (see Figure \ref{histo_plot}(c2)-(e2)). In particular, for both tumour types, the number of immunogenic cells tends to remain stable or decreases slightly. On the other hand, the number of non-immunogenic cells increases and grows to up to twice its initial value.\\
These results suggest that, beyond a certain non-immunogenic cell percentage threshold, the immune system becomes inefficient in both tumour types independently of the number of sub-populations of tumour cells. Moreover, they suggest that the selective pressure of the immune response can lead to more aggressive tumours, characterised by larger percentages  of non-immunogenic cells. In this regard, our results follow the same behaviour of previous experimental works demonstrating that, under cancer therapeutics (\textit{e.g.} chemotherapy or radiotherapy), the population of tumour cells is exposed to the selective stress induced by the treatment \citep{greaves2012clonal, iwasa2006evolution, tredan2007drug}. Therefore, more resistant cells acquire a competitive advantage over more sensitive cells and induce a weaker response to treatment in the long run. The resulting outcome is a more aggressive tumour, which may ultimately grow again \citep{nowell1976clonal}.

\section{Conclusions and perspectives}
\label{Discussion and conclusion}
The number of sub-populations of cancer cells constituting  a tumour and the percentage of immunogenic and non-immunogenic cells within it are two major components of ITH, and play a key role in the immune response against solid tumours. Mathematical models make possible to asses
the influence of these two components of ITH on anti-tumour immunity in a controlled manner.
\\
In this work, we have presented a spatially explicit stochastic individual-based model of the interaction dynamics between CD8$^+$ T cells and tumour cells, and we have investigated how ITH affects the anti-tumour immune response. \\
Our numerical results show that the number of sub-populations of cancer cells constituting a tumour can have a crucial impact upon the outcome of the immune response (Figures \ref{differentPopulations} and \ref{population}). In the scenario of tumours characterised by a low number of sub-populations of cancer cells, immune clearance can occur. Conversely, tumours composed of a larger number of sub-populations of cancer cells may be able to escape immune recognition and ultimately grow again. Our results suggest that increasing the number of sub-populations of cancer cells reduces the exposition of each antigen to the ``front-line'', thus making more difficult for the immune cells to detect them. (Figure \ref{differentPopulations}). Moreover, when tumours expressing clonal and sub-clonal antigens are considered, our results demonstrate that, in more heterogeneous tumours, tumour cells could have a better chance of escaping immune surveillance. This outcome may be explained by the fact that clonal antigens undergo ``dilution'' within the tumour relative to other antigens, diminishing the ability of CD8$^+$ T cells to mount a sufficient cytotoxic response (Figure \ref{population}). \\
The outcomes of our model support the idea that varying the initial percentage of immunogenic and non-immunogenic cells leads to variations on  the effectiveness of the immune response and results in distinct scenarios, from immune clearance of the tumour to immune escape (Figure \ref{boxplot_homogeneous}). 
We have also observed that for certain intermediate percentages  of immunogenic and non-immunogenic cells, stochasticity in cell dynamics plays an important role, and can lead both scenarios close to tumour eradication and to scenarios where a large number of tumour cells persists over time (Figure \ref{33_stoch}). 
\\
We have also studied the effects of ITH on anti-tumour immune response by varying both the number of sub-populations of cancer cells  and the initial percentage of immunogenic and non-immunogenic cells (Figure \ref{histo_plot}). For equal percentage of immunogenic and non-immunogenic cells, tumours with increased number of sub-populations of cancer cells are more aggressive than tumours with lower number of sub-populations of cancer cells. However, beyond a certain threshold value of the percentage of non-immunogenic cells, the immune system becomes inefficient against both types of tumours, independently of the number of sub-populations of cancer cells. In addition, we found that increasing initial percentages  of non-immunogenic cells always led to a less effective CD8$^+$ T cell response. When the tumours are not eradicated, the final percentage of non-immunogenic cells is larger than the initial one. This suggests that the immune system may act as a bottleneck which selects and eliminates immunogenic cells, thus allowing the tumours to escape immune regulation.  \\
In summary, our findings demonstrate the importance of ITH as a possible predictor of the outcome of immune action. Our results support the idea that patients with tumours bearing few clonal antigens are expected to be more likely to exhibit a durable benefit from immune response than patients with heterogeneous tumours characterised by many different sub-clonal antigens \citep{mcgranahan2016clonal}. On the other hand, our results disbelieve the fact that highly heterogeneous tumours, characterised by the expression of many different antigens, can enhance the efficacy of immune response. In fact, our results indicate that excessive antigen heterogeneity may, conversely, actively impair anti-tumour CD8$^+$ T cell immune response. This is also supported by a recent clinical work which found that excessive mutagenesis, directed to enhance the tumour mutational burden, may decrease the efficacy of immunotherapy \citep{wolf2019uvb}. \\

The current version of our model can be developed further in several ways. We could incorporate extended aspects of the tumour micro-environment, such as the expression of immunosuppressive factors (\textit{e.g.} PD1 or CTLA4), which affect the effectiveness of anti-tumour immune response. In fact, these inhibitory factors induce the exhaustion of CD8$^+$ T cells in the tumour micro-environment impairing the immune response \citep{jiang2015t,wherry2011t}. The inclusion of CD8$^+$ T cell exhaustion caused by inhibitory factors could give further explanations for other mechanisms of immune escape. The exhaustion mechanism could be included in the model by, for example, altering the value of the parameter governing the efficiency of the CD8$^+$ T cell population in eliminating tumour cells.  \\
The spatial dimension and the flexibility of our model would also allow for the study of the spatial distribution of CD8$^+$ T cells within the tumour and the role of immune infiltration on the tumour dynamics \citep{galon2019approaches}. Moreover, by posing the model on a 3D domain, a deeper understanding of the spatial dynamics of tumour-immune interactions could be achieved. \\
We managed to estimate some parameters of the model (see Table \eqref{table1} and Table \eqref{table2}) from the literature and define them on the basis on precise biological assumptions. However, there are some parameters (\textit{e.g} $r, C_1, C_2$ and the parameters related to the chemoattractants) whose values were simply chosen with an exploratory aim and to qualitatively reproduce essential aspects of the experimental results obtained in \citep{wolf2019uvb}. In order to minimise the impact of this limitation on the conclusions of our study, we carried out simulations by keeping all parameter values fixed and changing only the initial composition of the tumours, and then comparing the simulation results so obtained.\\
Finally, from a modelling point of view, although more tailored to capture fine details of the dynamics of single cells, individual-based models are not amenable to analytical studies, which may support a more in-depth theoretical understanding of the application problems under study. For this reason, in future work we plan to derive a continuum model from a simplified version of our individual-based model by using mean filed methods similar to those employed in \citep{ardavseva2020comparative,chisholm2016evolutionary, painter2015navigating}. \\
In its present form, our modelling framework qualitatively reproduces scenarios of successful and unsuccessful immune surveillance reported in experimental studies \citep{gejman2018rejection,mcgranahan2016clonal} and \citep{wolf2019uvb}. However, at this stage, the model has not been calibrated using any particular type of data. Hence, it cannot be employed to generate predictions that can directly be used in the clinic. By fitting its parameters to a specific type of clinical data, our model could, in principle, be used to assess different levels of ITH as potential biomarkers for comparing and predicting outcomes in tumour immunotherapy treatments. Integrating the model with tumour biopsies from patients could offer insight into potential outcomes of treatments. Finally, our model may be a promising tool to explore therapeutic strategies designed to decrease tumour heterogeneity and improve the overall anti-tumour immune response. 

\section*{CRediT authorship contribution statement} 
\textbf{Emma Leschiera}: Conceptualization, Methodology,  Software, Formal analysis, Investigation, Visualization, Writing - original draft. \textbf{Tommaso Lorenzi}: Conceptualization, Methodology, Writing - Review $\&$ Editing, Supervision. \textbf{Shensi Shen}: Conceptualization, Methodology, Resources, Writing - Review $\&$ Editing. \textbf{Luis Almeida}: Conceptualization, Methodology, Writing - Review $\&$ Editing, Supervision. \textbf{Chloe Audebert}: Conceptualization, Methodology, Formal analysis, Validation, Visualization, Writing - original draft, Supervision. 

\section*{Acknowledgments}
The authors are grateful to Jacqueline Marvel for sharing her knowledge on the immune response. \\
E.L. has received funding from the European Research Council (ERC) under the European Union’s Horizon2020 research and innovation programme (grant agreement No 740623). \\
T.L. gratefully acknowledges support of the MIUR grant ``Dipartimenti di Eccellenza 2018-2022''.
\appendix
\section{Details of computational model}
\label{Details of computational model}
The individual-based model has been numerically solved using the multicellular modelling environment CompuCell3D \citep{izaguirre2004compucell}. This software is an open source solver, which uses a Cellular Potts model \citep{graner1992simulation} (also known as CPM, or Glazier-Graner-Hogeweg model). In Cellular Potts models, biological cells are treated as discrete entities represented as a set of lattice sites, each with characteristic values of area, perimeter, and intrinsic motility on a regular lattice. Interaction descriptions and dynamics between cells are modelled by means of the effective energy of the system. This determines many characteristics such as cell size, motility, adhesion strength and the reaction to gradients of chemotactic fields. During a simulation, each cell will attempt to extend its boundaries, through a series of index-copy attempts, in order to minimise the effective energy. The success of the index copy attempt is dependent upon probabilistic rules which take into account the change in energy. 
\subsection{Cell types}
In Cellular Potts models, cells are uniquely identified with an index $\sigma(i)$ on each lattice site $i$, with $i$ a vector of integers occupying lattice site $i$. Each cell in the model has a type $\tau(\sigma(i))$, which determines its properties, and the processes and interactions in which it participate.
In our model, to characterise the different sub-populations of cancer cells, we define as many types of tumour cells as sub-populations of cancer cells. Moreover, CD8$^+$ T cells have as many types as TCRs considered for the simulation, which corresponds also to the number of antigens considered. Note
that, technically, the extracellular medium is also considered as a cell of type medium.
\subsection{Cellular dynamics}
The effective energy is the basis for operation of all Cellular Potts models, including CompuCell3D \citep{izaguirre2004compucell}, because it determines the interactions between cells (including the extracellular medium). Configurations evolve to minimise the effective energy $H$ of the system, defined as

\begin{equation}
\footnotesize
\begin{aligned}
H&=\underbrace{\sum_{i,j}J(\tau(\sigma_i),\tau(\sigma_j))(1-\delta(\sigma_i,\sigma_j))}_{\text{boundary energy}}+\underbrace{\sum_\sigma\left[\lambda_{area}(\sigma)(a(\sigma)-A_t(\sigma))^2\right]}_{\text{area constraint}}+\underbrace{\sum_\sigma\left[\lambda_{per}(\sigma)(p(\sigma)-P_t(\sigma))^2\right]}_{\text{perimeter constraint}}
\label{Hsum}
\end{aligned}
\end{equation}
The most important component of the effective energy equation is the boundary energy, which governs the adhesion of cells. The boundary energy $J(\tau(\sigma_i),\tau(\sigma_j))$ describes the contact energy between two cells $\sigma_i$ and $\sigma_j$ of types $\tau(\sigma_i)$ and $\tau(\sigma_j)$. It is calculated by the sum over all neighbouring pixels $i$ and $j$ that form the boundary between two cells. Thanks to the term $(1-\delta(\sigma_i,\sigma_j))$, the boundary energy contribution is considered only between lattice sites belonging to two different cells. The second and third terms represent respectively a cell-area and cell-perimeter constraint. In particular, $a(\sigma)$ and $p(\sigma)$ are the surface area and perimeter of the cell $\sigma$, $A_t(\sigma)$ and $P_t(\sigma)$ are the cell’s target surface area and perimeter and $\lambda_{area}(\sigma)$ and $\lambda_{per}(\sigma)$ are an area and perimeter constraint coefficient. 

The cell configuration evolves through lattice-site copy attempts. To begin an index-copy attempt, the algorithm randomly selects a lattice site to be a target pixel $i$, and a neighbouring lattice site to be a source pixel $i^\prime$. If the source and target pixels belong to the same cell (\textit{i.e.} if $\sigma(i) =\sigma(i^\prime))$, they do not need to attempt an lattice-site copy and thus the effective energy will not be calculated. Otherwise, an attempt will be made to switch the target pixel as the source pixel, thereby increasing the surface area of the source cell and decreasing the surface area of the target cell.  \\
The algorithm computes $\Delta H=H-H^\prime$, with $H$ the effective energy of the system and $H^\prime$ the effective energy if the copy occurs. Then, it sets $\sigma(i) =\sigma(i^ \prime)$ with probability $P(\sigma(i)\rightarrow\sigma(i^\prime))$, given by: 
\begin{equation}
    P(\sigma(i)\rightarrow\sigma(i^\prime))= \begin{cases}\quad 1 \; \;  \quad : \quad \Delta H \le 0 \\
    \exp^{-\frac{\Delta H}{T_m}} \; : \quad \Delta H > 0.
    \end{cases}
    \label{Boltzmann}
\end{equation}
 The change in effective energy $\Delta H$ evaluate the energy cost of such a copy and parameter $T_m$ determines the level of stochasticity of accepted copy attempts. The
unit of simulation time is the Monte Carlo step (MCS).
\subsection{Subcellular dynamics and chemotaxis}
 In our model we simulate CD8$^+$ T cell chemotaxis toward tumour cells, defined as the cell motion induced by a presence of a chemical. In CompuCell3D \citep{izaguirre2004compucell}, chemotaxis is obtained biasing the cell’s motion up or down a field gradient by adding a term $\Delta H_{chem}$ in the calculated effective-energy change $\Delta H $ used in the acceptance function \eqref{Boltzmann}. For a field $c(i)$: 
\begin{equation}
    \Delta H_{chem}=-\lambda_{chem}(c(i)-c(i^\prime))
    \label{chem}
\end{equation}
where $c(i)$ is the chemical field at the index-copy target pixel $i$, $c(i^\prime)$ the field at the index-copy source pixel $i^\prime$, and $\lambda_{chem}$ the strength and direction of chemotaxis. \\
The change in concentration of the chemical field $c$ is obtained by solving a reaction-diffusion equation of the following general form: 
\begin{equation}
  \frac{\partial c}{\partial t}= D\nabla^2c-\gamma c+S  
\end{equation}
where $D$, $\gamma$ and $S$ denote the diffusion constant, decay constant and secretion rates of the field, respectively. These three parameters may vary with position and cell-lattice configuration, and thus be a function of cell $\sigma$ and pixel $i$.
\section{Model parameters}
\label{appendix:graph}

The individual-based model is parametrised using parameter values obtained from published biological data wherever possible. We use a 2D squared spatial domain with $400\times 400$ lattice sites (pixels). We assume that a pixel of the domain corresponds to 3 $\times$ 3 $\mu m^2$. As the CD8$^+$ T cell diameter is estimated to be between $10$ $\mu m$ and $12$ $\mu m$ \citep{gao20162,gong2017computational}, the initial size of a CD8$^+$ T cell is $4\times4$ pixels. A tumour cell diameter is estimated to be about 20 $\mu m$ \citep{christophe2015biased}, therefore we assume that each
newly divided tumour cell is made of $5\times 5$ pixels.  In addition, the maximum CD8$^+$ T cell migration speed measured in the simulation is around 10 pixels / 100 MCS. Therefore, using the CD8$^+$ T cell migration measurements \textit{in vivo} (2-25 $\mu m/min$, see \citet{miller2003autonomous}) we choose 1 MCS $\sim$ 1 minute as a time scale.
The parameters for the Cellular Potts model are listed in Table \ref{table1}, while all the other parameters with their related references are listed in Table \ref{table2}. \\
We now provide a discussion on how some of the parameters of the Cellular Potts model were chosen. Interactions between neighboring pixels in the Cellular Potts model have an effective energy, $J$ (as it appears in Equation \eqref{Hsum}), which characterises the strength of cell-cell adhesion (see Table \ref{table1}). A larger $J$ means that more energy is associated with the interface between two cells, which is less energetically favourable, corresponding to weaker adhesivity. It can be observed that $J_{CT}$
 and $J_{TT}$
 are lower than $J_{CC}$. We make the assumption that tumour cells stay in contact to compactly create the tumour mass and that when a CD8$^+$ T cell enters in contact with a tumour cell then strongly binds to it. When they migrate in the domain to search for tumour cells, CD8$^+$ T cells are not in contact with each other.\\
 Files to run a simulation example with Compucell3D software are available at: \url{https://plmlab.math.cnrs.fr/audebert/cc3dmodeltumourcd8}.
 
\begin{table}[hp!]
	\small
\caption{Parameter values used to implement the Cellular Potts model. Energies, temperature and constrains are dimensionless parameters.}
\centering
 	\begin{tabular}{p{1.5cm} p{1.2cm} p{7cm} p{2.8cm} p{1.5cm}}
  \hline
  Phenotype & Symbol & Description & Value & Reference \\
  \hline
  \textbf{Domain} & $\Delta x, \Delta y $ & Domain spacing in the $x$ or $y$ direction& $1$ Pixel $=3\times3\; \mu m^2$ &  \\ 
   & $\Delta t$ &Time-step&  1 MCS = 1 min & \\
     & $t_f$ & Final time &  $20$ (\textit{days})  & \\ \\
       \textbf{CC3D} & $J_{MT}$ & Contact energy tumour cells-medium & 50 &  \\ 
       & $J_{MC}$ & Contact energy CD8$^+$ T cells-medium & 50 &  \\ 
       & $J_{CT}$ & Contact energy CD8$^+$ T cells-tumour cells & 20 &  \\ 
       & $J_{TT}$ & Contact energy tumour cells-tumour cells & 110 &  \\ 
       & $J_{CC}$ & Contact energy CD8$^+$ T cells-CD8$^+$ T cells & 1000 &  \\
       & $d_{T}$ & Tumour cell diameter  & 20-40 $(\mu m)$ & \citep{gong2017computational}\\
       & $d_{C}$ & CD8$^+$ cell diameter & 12 $(\mu m)$ &  \citep{gao20162}\\
       & $\lambda_{area}$ & Tumour cell and CD8$^+$ T cell area constrain  & $10$ & \\
       & $\lambda_{per}$ & Tumour cell and CD8$^+$ T cell perimeter constrain  & $10$ & \\
        & $T_m$ & Fluctuation amplitude parameter & $10$ & \\
        & $\lambda_{chem}$ & Strength and direction of chemotaxis  & $50$ & \\
  \hline
\end{tabular}
\label{table1}
\end{table}

\begin{table}[hp!]
	\footnotesize
\caption{Parameter values used in numerical simulations.}
\centering
 	\begin{tabular}{p{2.1cm} p{4.9cm} p{5.8cm} p{1.5cm}}
  \hline
  Phenotype  & Description & Value & Reference \\
  \hline
  \textbf{Tumour} & Initial number  &$N_T(0) =400$ &  \\ 
      & Index identifier& $n=1, \dots, N_T(t)$&  \\
  
   &  Lifespan   & $\mathcal{U}_{[3,7]}$ (\textit{days}) &  \citep{gong2017computational}\\
   
    & Growth rate  & $\mathcal{U}_{[0.015,0.02]}$ (\textit{pixel/min}) &   \citep{gong2017computational}
 \\

  & Mean cycle time  & $24$ (\textit{hours}) &  \citep{gong2017computational}  
 \\
  & Rate of death due to competition between tumour cells  & $3.8\times 10^{-7}$ (\textit{1/min}) &   
 \\ & Antigen profile  & $A=(a_1,\dots,a_f)$&   
 \\
  & Clonal antigen profile  & $A_{C}\subset A$&   
 \\
  & Sub-clonal antigen profile  & $A_{SC}\subset A$&   
 \\
  & Range of value of antigen presentation for clonal cells & $L_{I}=\left\{\frac{1}{6},\frac{2}{6},\frac{3}{6},\frac{4}{6},\frac{5}{6},1\right\} $ \\
   & Range of value of antigen presentation for sub-clonal cells & $L_{NI}=\left\{\frac{5}{100},\frac{10}{100},\frac{15}{100},\frac{20}{100},\frac{25}{100},\frac{30}{100}\right\} $ \\
  & Antigen  & $a_i \in A$&   
 \\
  & Level of presentation of antigen $a_i$ of tumour cell $n$  & $l_{a_i}^n \in [0,1]$&   
 \\\\

  \textbf{CD8$^+$ T cells}  & Total number at time $t$ & $N_C(t) \geq 0 $&  \\
  & Index identifier  & $m=1, \dots, N_C(t)$&  \\
  & Influx rate & $p(t)=C_1\times S_a^{tot}(t)$   &   \\ 
     &  & $C_1=2\times 10^{-5}$ (\textit{min/mol})&   \\
   & Lifespan & $\mathcal{U}_{[2.5,3.5]}$ (\textit{days})&  \citep{gong2017computational}\\
   & Migration speed & 2 - 25 (\textit{$\mu m$/min}) & \citep{gao20162}\\
    & Engagement time & 6 (\textit{hours})  &  \citep{christophe2015biased} \\
     & TCR-recognition probability  & 	$r=0.901$  &   \\
      \\
     \textbf{Chemoattractant} 
    & Concentration &  $c_{a_i}\geq 0 $ (\textit{mol/pixel}) &  \\
     & Total amount secreted &  $S_{a_i}^{tot}\geq 0 $ (\textit{mol/min}) &  \\
     & Diffusion & $ D=7\times 10^{-1}$ (\textit{pixel$^2$/min}) &  \\
       & Secretion & $s_{a_i}^n=C_2 \times l_a^n$ (\textit{mol/min/pixel})\\
       &  & $C_2=10$ (\textit{mol/min/pixel}) \\
       & Decay & $\gamma=3\times 10^{-4} $ (\textit{1/min}) &  \\
       & Initial concentration & $c_{a_i}^{init}=0.5(280-\sqrt{(x-200)^2+(y-200)^2})$ &  \\
  \hline
\end{tabular}
\label{table2}
\end{table}

\newpage



\bibliography{bibliography.bib} 
\end{document}